\documentclass[fleqn,usenatbib]{mnras}

\usepackage{newtxtext,newtxmath}

\usepackage[T1]{fontenc}

\DeclareRobustCommand{\VAN}[3]{#2}
\let\VANthebibliography\thebibliography
\def\thebibliography{\DeclareRobustCommand{\VAN}[3]{##3}\VANthebibliography}

\usepackage{graphicx}	
\usepackage{amsmath}	
\usepackage{adjustbox}  
\usepackage[normalem]{ulem}

\title[JWST CEERS MIRI source counts]{Source counts at 7.7 to 21 $\mu$m in CEERS field with James Webb Space Telescope}

\author[Wu et al. 2022]{
Cossas K.-W. Wu$^{1}$,
Chih-Teng Ling$^{2}$,
Tomotsugu Goto$^{1,2}$, 
Seong Jin Kim$^{2}$,
Tetsuya Hashimoto$^{4}$,
\newauthor
Ece Kilerci$^{3}$,
Yu-Wei Lin$^{1,2}$,
Po-Ya Wang$^{1}$,
Yuri Uno$^{4}$,
Simon C.-C. Ho$^{2,5}$, and 
\newauthor
Tiger Yu-Yang Hsiao$^{2}$
\\
$^{1}$Department of Physics, National Tsing Hua University, 101, Section 2. Kuang-Fu Road, Hsinchu, 30013, Taiwan (R.O.C.)\\
$^{2}$Institute of Astronomy, National Tsing Hua University, 101, Section 2. Kuang-Fu Road, Hsinchu, 30013, Taiwan (R.O.C.)\\
$^{3}$Sabanc{\i} University, Faculty of Engineering and Natural Sciences, 34956, Istanbul, Turkey\\
$^{4}$Department of Physics, National Chung Hsing University, 145, Xingda Road, Taichung, 40227, Taiwan (R.O.C.)\\
$^{5}$Research School of Astronomy and Astrophysics, The Australian National University, Canberra, ACT 2611, Australia\\
}

\date{Accepted 2023 June 8. Received 2023 June 1; in original form 2022 September 5}

\pubyear{2022}

\begin{document}
\label{firstpage}
\pagerange{\pageref{firstpage}--\pageref{lastpage}}
\maketitle

\begin{abstract}

Source counts --- the number density of sources as a function of flux density --- {represent one of the fundamental metrics in observational cosmology due to their straightforward and simple nature}. {It is an important tool that provides information on galaxy formation and evolution. Source counting is a direct measurement. Compared to advanced analyzes that require more observational input such as luminosity/mass functions, it is less affected by any cosmological parameter assumptions or any errors propagated from luminosities.} In this {study}, we present source counts at the six mid-infrared bands, i.e., 7.7, 10, 12.8, 15, 18, and 21 $\mu$m from the MIR instrument of the James Webb Space Telescope (\textit{JWST}). {Contrasted with the infrared source counts achieved by prior generations of infrared space telescopes, our source counts delve up to $\sim$100 times deeper, showcasing the exceptional sensitivity of the \textit{JWST}, and aligning with the model predictions based on preceding observations. In a follow-up study, we utilize our source counts to establish a new IR galaxy population evolutionary model that provides a physical interpretation.}
\end{abstract} 



\section{Introduction}
With the successful launch of the James Webb Space Telescope \citep[\textit{JWST}]{Gardner2006, Kalirai2018} in 2021, observational cosmology {has started} to benefit from unprecedented sensitivity gains in the infrared (IR). {This highly advanced} telescope is now leading us to explore the faintest IR populations not only near our Galaxy but also in the distant Universe. Previous space missions for IR observations, such as the Infrared Astronomical Satellite \citep[\textit{IRAS},][]{Neugebauer1984}, \textit{ISO} \citep{Kessler1996}, \textit{AKARI} \citep{Murakami2007}, \textit{Spitzer} \citep{Werner2004}, and \textit{Herschel} \citep{Pilbratt2010} have revealed a variety of IR populations and their evolutionary properties, {as seen through} the luminosity functions (LFs) \citep[e.g.,][]{Saunders1990, Rowan-Robinson1997, Elbaz2002, Caputi2007, Goto2010, Gruppioni2010, Gruppioni2011}. {Additionally}, in the era of \textit{Spitzer}, deep-field surveys were carried out to analyze the faint IR populations. In the extreme deep field, e.g., S-CANDELS, SEDS, \citet{Ashby2015} {has} already reached sub-uJy levels at 4.5$\mu$m. The \textit{JWST} {is set to} continue and extend {the work of} its precursors to {longer} wavelengths and even more sensitive deep field surveys.

Astronomers rely on dusty and star-bursting galaxies to explain the observed cosmic infrared background. These galaxies re-radiate a substantial amount of their bolometric energy, primarily in the IR wavelength. {In particular}, the spectral energy distribution (SEDs) of mid-IR (MIR) galaxies can identify emissions from star-forming (SF) activities and active galactic nuclei (AGNs, including type 1 and 2). {Based on the} emission {characteristics} of galaxies, they can be broadly classified into four categories: star-forming (SF) galaxies, AGNs, a mixture of the two, and quiescent galaxies. SF and AGN characteristics {are thought to be} key to the evolving population of source counts in the MIR. {Therefore}, observing the number density of both AGNs and SF galaxies {can help decipher the critical aspects of} cosmic galaxy evolution history.

The source counts of bright, IR galaxies have been obtained in ISO bands \citep[e.g.,][]{Pearson2005}, Spitzer bands \citep[e.g.,][]{Pearson2005}, and AKARI bands \citep[e.g.,][]{Wada2008, Pearson2010, Pearson2014, Takagi2012}. {Nevertheless, at} the faint end of the MIR source counts, luminosity functions (LFs), and evolution models reaching down to the sub-$\mu$Jy levels {remain partially unexplored} in the pre-\textit{JWST} era, {with these aspects having been} only predicted by several studies \citep[e.g.,][]{Gruppioni2011, Cowley2018}.

The latest source counts using \textit{JWST} data \citep[][]{Ling2022} focused on the extra fields surrounding Stephan's Quintet, which {are} less contaminated by foreground objects such as Galactic stars. However, due to the lack of IR spectral observations from \textit{JWST} in the current Stephan's Quintet field, \citet{Ling2022} were unable to provide any quantitative remarks on potential contamination/noise from the foreground. \citet{Ling2022} had to manually mask out and exclude contaminants or unnecessary objects/regions from their work, potentially {resulting in a shallower} flux limit. Compared to Stephan's Quintet field studied by \citet{Ling2022}, the Cosmic Evolution Early Release Science Survey (Finkelstein et al. in prep., hereafter CEERS) provides a much {broader and cleaner} field for extra-galactic studies, suffering less from the foreground contaminants. Therefore, we {utilize} a clean field to obtain continuous MIR source counts with CEERS, hoping that it will provide us with crucial early insights into the evolutionary properties of MIR-selected galaxies.

In this work, we only focus on the {two pointings} having full MIRI coverage from 7.7 to 21.0 $\mu$m of the first early MIRI data release on Jul. 14, 2022.
The CEERS survey will cover about 100 sq. arcmin in the future, taking some parts of the Extended Groth Strip (EGS) field using NIRCam, MIRI, and NIRSpec.
More details of the observational strategy can be found on the official website of CEERS Survey \footnote[1]{\url{https://ceers.github.io/obs.html}}. 

This paper is organized as follows: 
In \ref{S:data} we present the basic {properties} of \textit{JWST} CEERS data, source extraction, and the completeness of our source detection.
In \ref{S:res}, we discuss our observed source count results with the model predictions from the literature.
Finally, our conclusion is given in \S \ref{S:conc}.
We follow \citet{Cowley2018}, whose models we compare with, adopting the {\it Planck15} cosmology \citep{Planck2016}, i.e., $\Lambda$ cold dark matter cosmology with ($\Omega_{m}$, $\Omega_{\Lambda}$, $\Omega_{b}$,$h$)=(0.307, 0.693, 0.0486, 0.677). 

\section{Data Analysis}
\label{S:data}
\begin{figure}
    \centering
    \adjincludegraphics[trim={{0.72\columnwidth} 0 0 0}, clip, width=0.7\columnwidth]{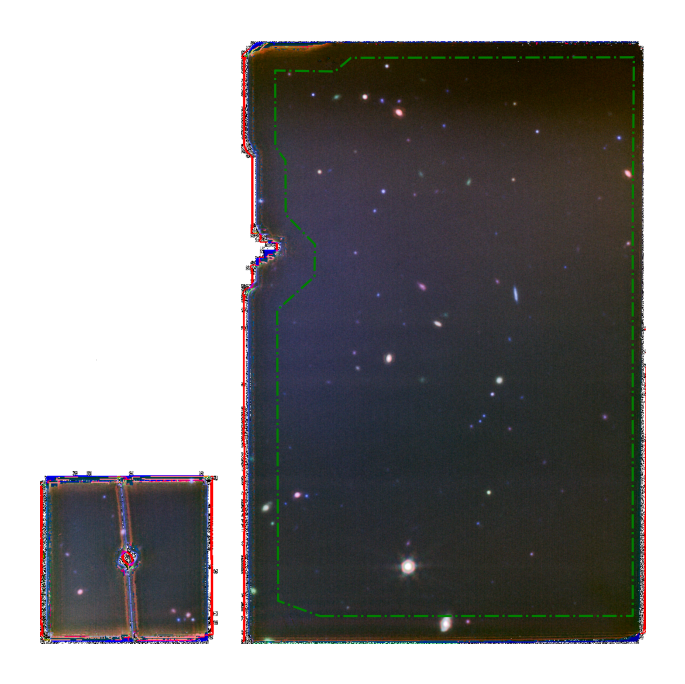}
    \adjincludegraphics[trim={{0.73\columnwidth} 0 0 0}, clip, width=0.7\columnwidth]{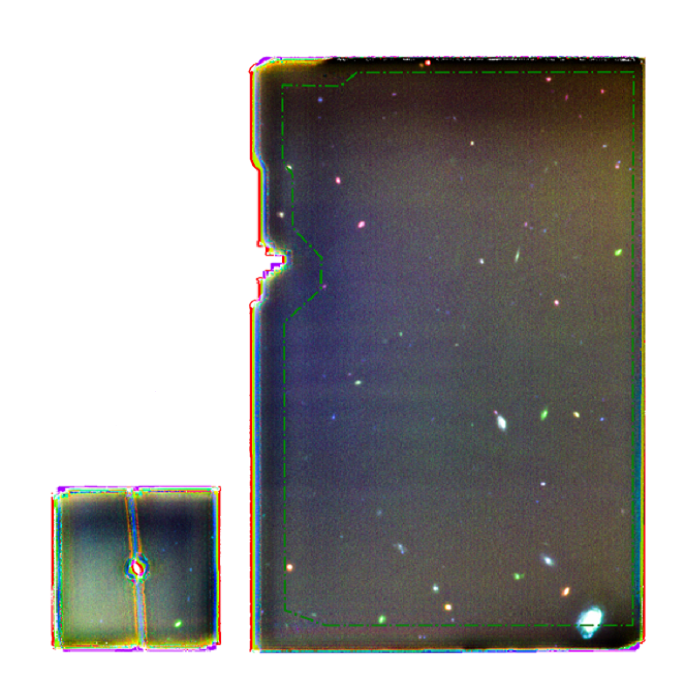}
    \caption{The composite false-colour image of CEERS with the MIRI 7.7-$\mu$m band (Blue), {the} 12.8-$\mu$m, band (Green) and {the} 18.0-$\mu$m band (Red). Upper: o001 field; lower: o002 field. Regions outside of the green dot-dashed polygon are not used in our analysis. The same mask is applied to the other filters as well.}
    \label{fig:1}
\end{figure}

We used data from one of the Early Release Science programs, dedicated to the Cosmic Evolution Early Release Science (CEERS; PID 1345) Survey \citep[][]{Finkelstein2017} by \textit{JWST}.
We did not perform any further map/flux calibration for the level-3 image products from the MAST archive.
The level-3 image products have gone through the JWST general calibration pipeline version 1.8.2, which was released on October 19th, 2022. Furthermore, as documented in the early calibration caveat, some of the MIRI images produced via early pipeline would still suffer from the so-called “shower artifacts” produced by cosmic rays. The artifact would affect the data quality and further impact the scientific analysis. Fortunately, shower artifacts are not {present} in the CEERS fields used in this work. Therefore, no additional major calibration is needed.
We selected the fields that are covered by all the broadband filters centered at 7.7 $\mu$m, 10.0 $\mu$m, 12.8 $\mu$m, 15.0 $\mu$m, 18.0 $\mu$m, and 21.0 $\mu$m as shown in Fig.\ref{fig:1}.
The IDs of the two fields are $\text{jw01345-o001\_t021}$ and $\text{jw01345-o002\_t022}$, respectively.  Hereafter, for simplicity and clarity, we use the notations 'o001' and 'o002' for $\text{jw01345-o001\_t021}$ and $\text{jw01345-o002\_t022}$, respectively.

\subsection{Source extraction \& Photometry} \label{SE:Photometry}

In each CEERS field, due to the higher {noise} along the edge of the images, we {masked} the edges from each image frame with the green polygon shown in Fig. \ref{fig:1} for data analysis in this work.
The noise mainly comes from the less dithering frames in the stacking procedure. 
The remaining areas in o001 and o002 field are 7127.45 arcsec$^2$ ($\sim 2$ arcmin$^2$), and 14254.9 arcsec$^2$ ($\sim 4$ arcmin$^2$) in total.
The pixel scale is 0.1109 arcsec/pixel in all filters.

While checking the public catalogue available on the JWST archive, we noticed that some sources close to the detection limits are missing. Since we aim to include faint sources in the MIR bands, we performed our own source extraction using the Source-Extractor V2.19.5 \citep[][hereafter SE]{Bertin1996}. We have followed the photometry of the previous work \cite{Ling2022} and improved it for accuracy with \textsc{Photutils}. \textsc{Photutils} is a Python package that could be used to handle basic astronomical image processing and analysis, which is also used in \textit{JWST} pipeline. Both \textsc{Photutils} \citep{Larry2022} and \textsc{SE} are involved in performing the photometry in this work, whereas \citet{Ling2022} relied only on \textsc{SE}.

In this work, \textsc{Photoutils} is used for background estimation and subtraction, and \textsc{SE} is used for source extraction, because we found that \textsc{Photoutils} performs better than \textsc{SE} in background estimation on these images. In \textsc{SE}, the background estimation sometimes erroneously includes the fluxes from the sources. This could lead to significant flux loss for the bright extended sources. However, \textsc{Photoutils} uses a smoother background estimation approach that does not confuse sources as a background. As a result, we perform background subtraction on each MIRI image using \textsc{Photoutils}.

\begin{figure}
    \centering
    \includegraphics[width=\columnwidth]{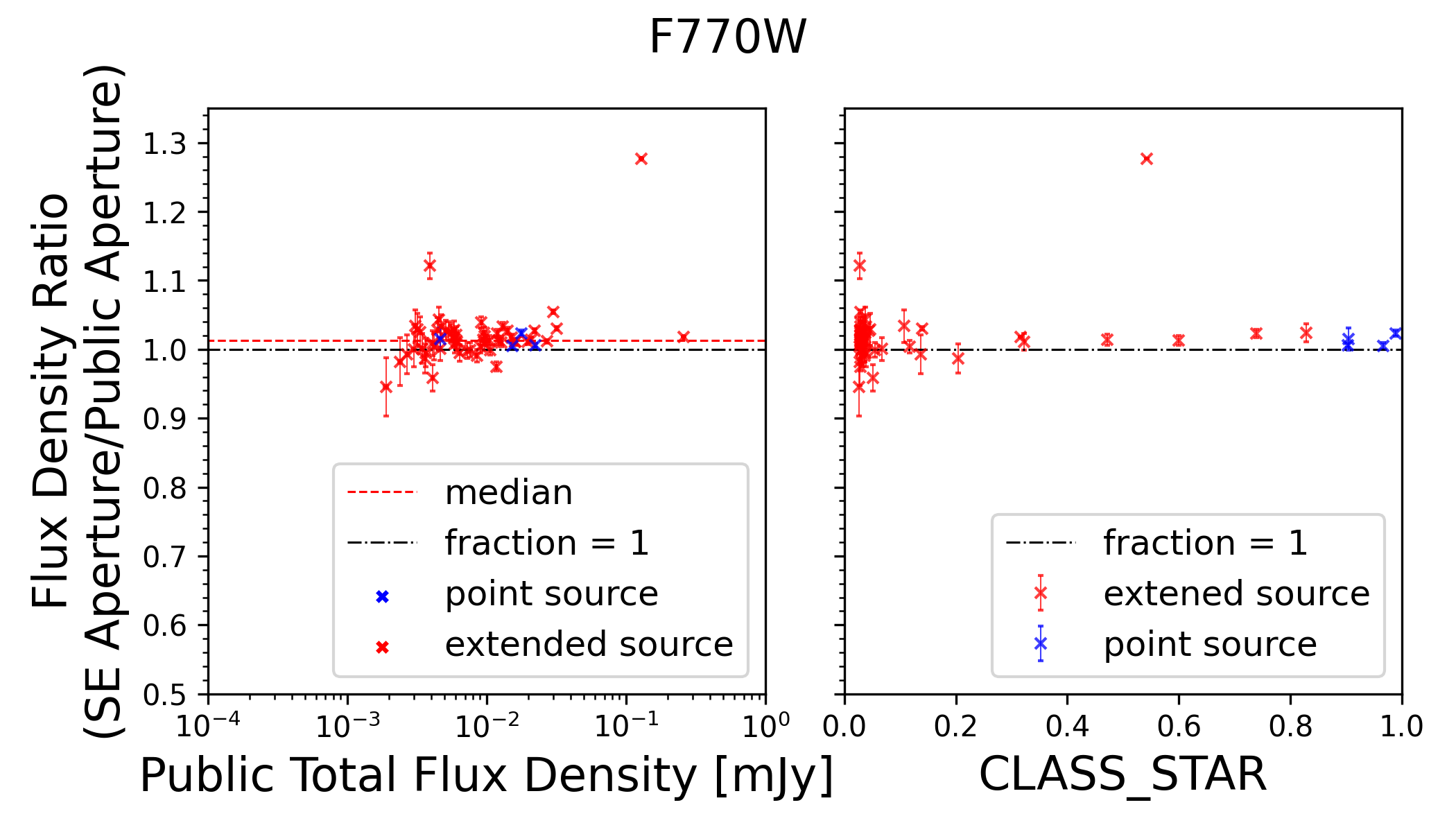}
    \includegraphics[width=\columnwidth]{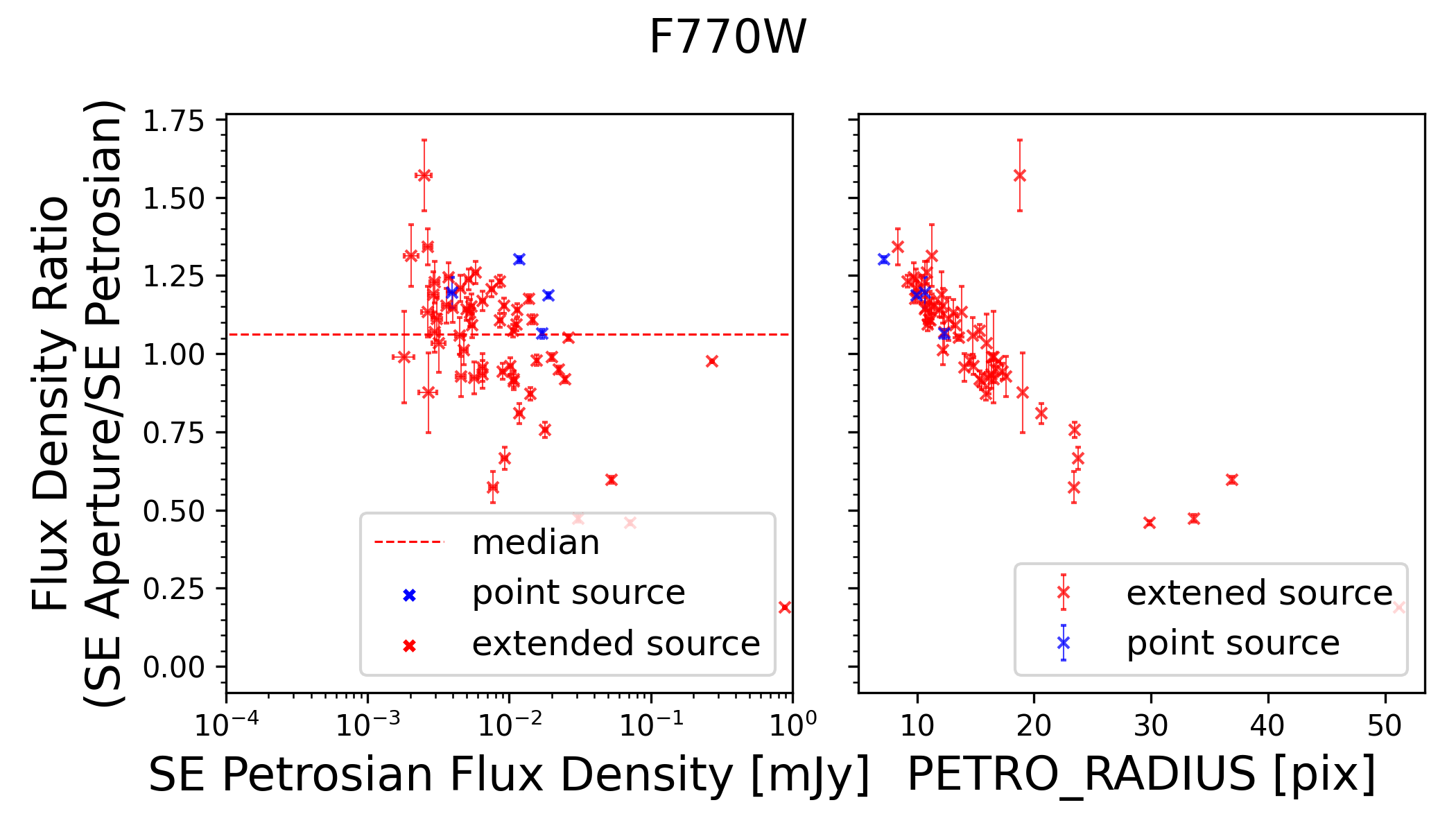}
    \caption{{{(upper)} The y-axis of the above panels shows the recovered flux from the \textsc{SE} divided by the flux measurements of the JWST public catalogs in the 7.7$\mu$m band. The ideal recovery would result in the black horizontal dashed-dotted at y=1 line. The red horizontal dashed lines indicate the median of the recovered flux among all sources. The 1-sigma deviation is 4\%, 12\%, 11\%, 10\%, 4\%, 7\% depending on the band in 7.7, 10, 12.8, 15, 18, 21 7.7$\mu$m, respectively. {(lower)} The y-axis represents the flux density ratio between fixed-size aperture photometry and Petrosian aperture photometry. In this work, Petrosian aperture photometry is applied to all the sources for a better measurement compared to the fixed-size aperture photometry in \citet{Ling2022}.}}
    \label{770ff}
\end{figure}

\begin{figure}
    \centering
    \includegraphics[width=\columnwidth]{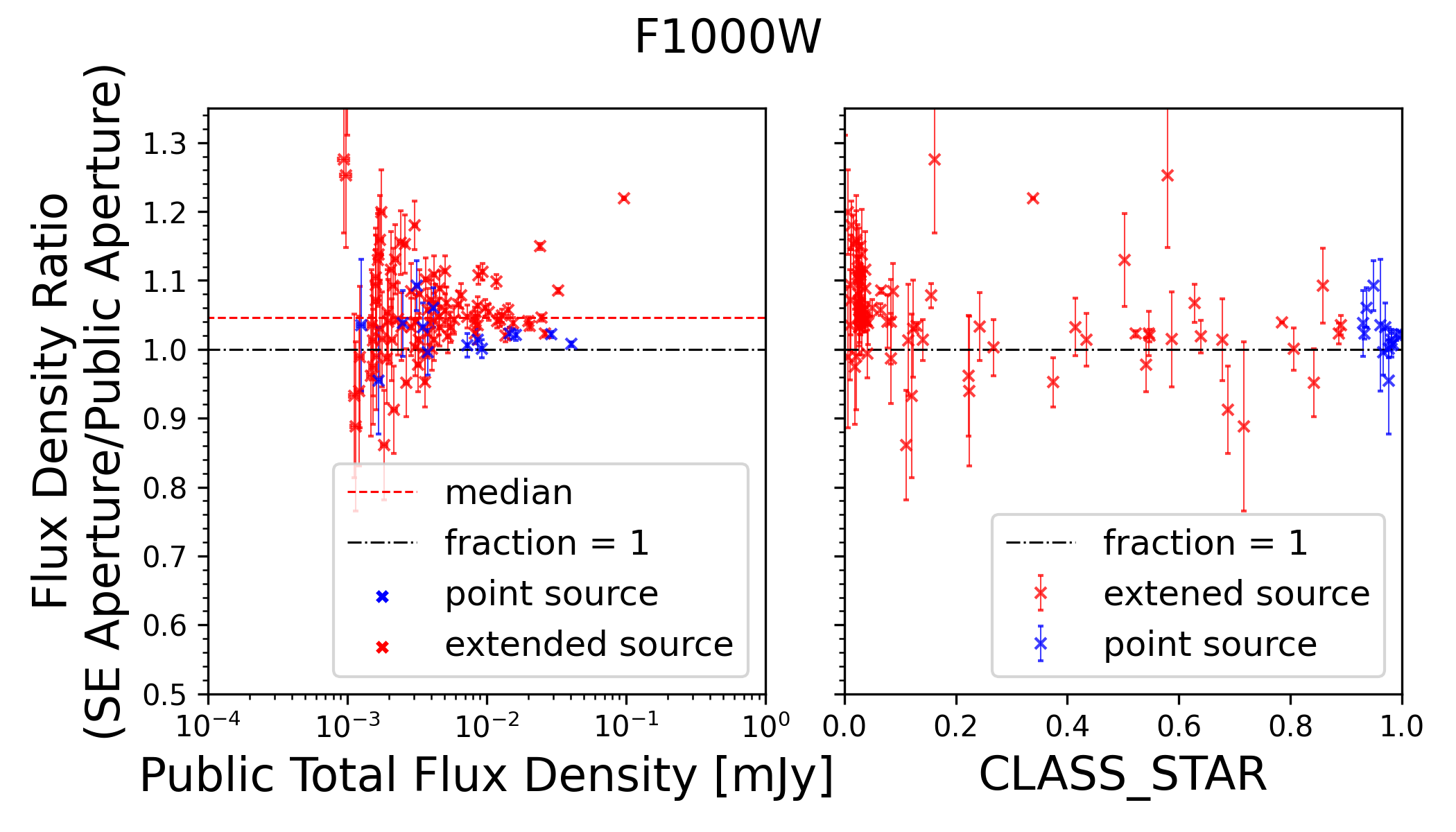}
    \includegraphics[width=\columnwidth]{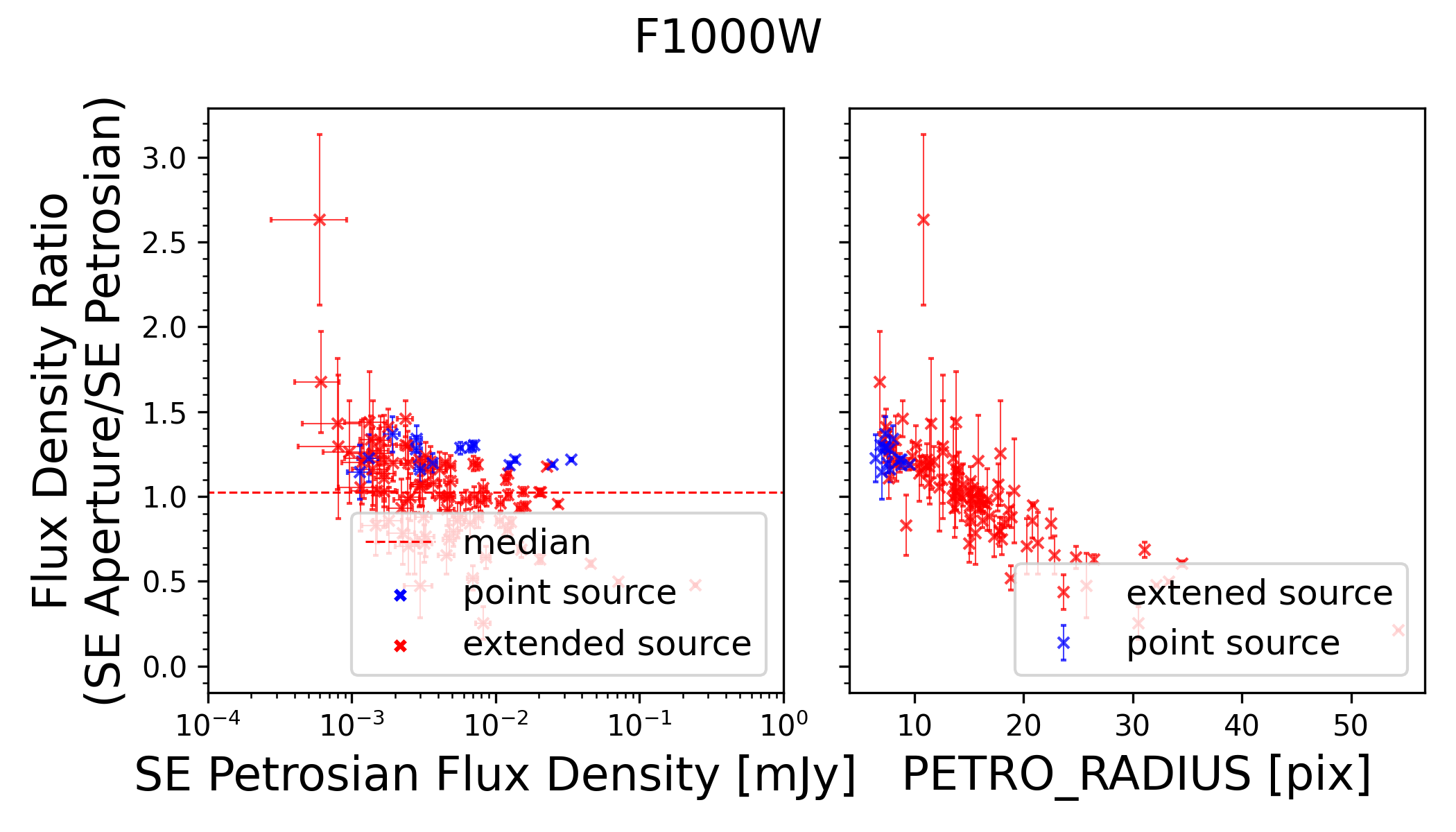}
    \caption{{Same as figure \ref{770ff} but for 10$\mu$m band.}}
    \label{1000ff}
\end{figure}

\begin{figure}
    \centering
    \includegraphics[width=\columnwidth]{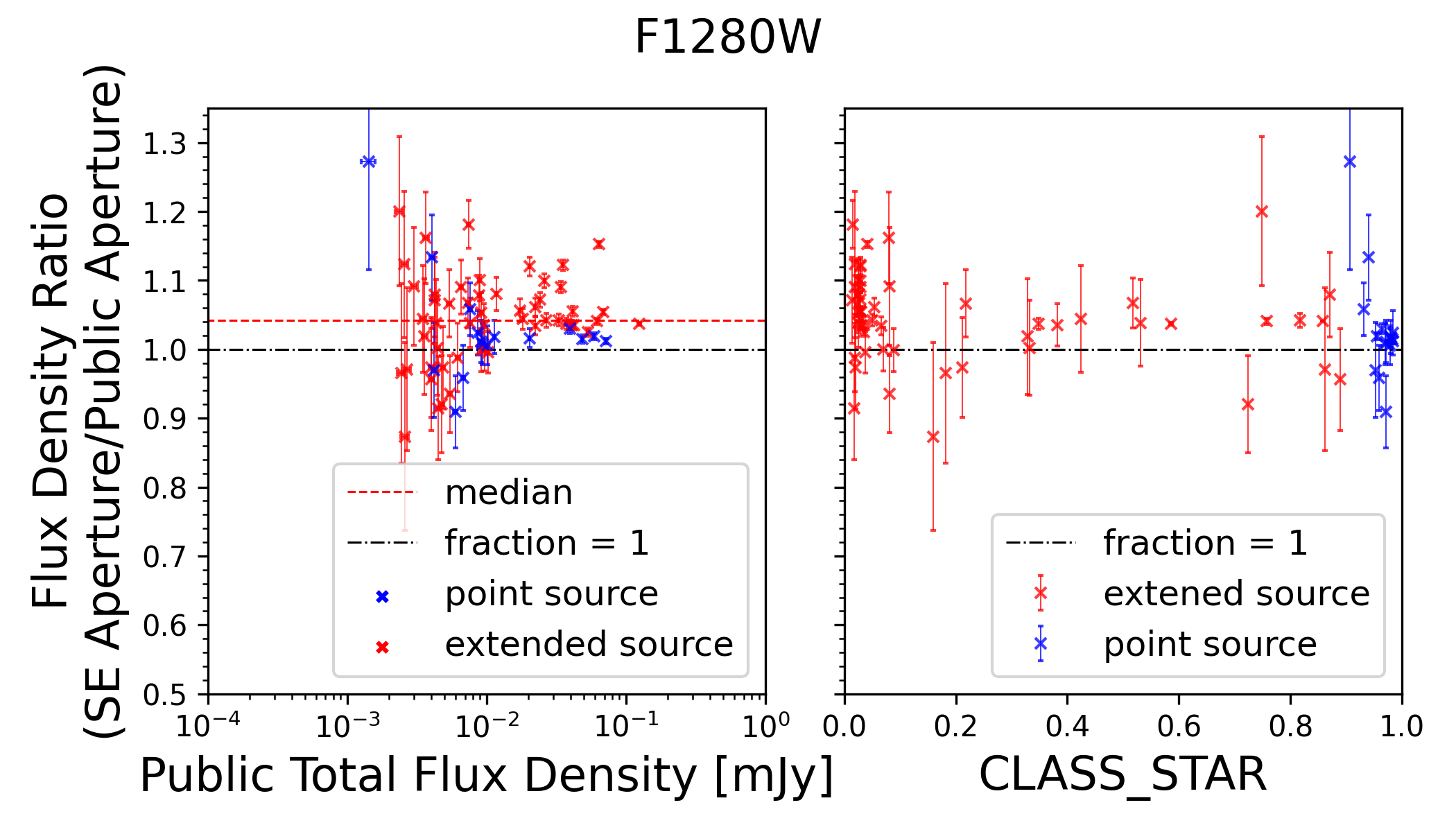}
    \includegraphics[width=\columnwidth]{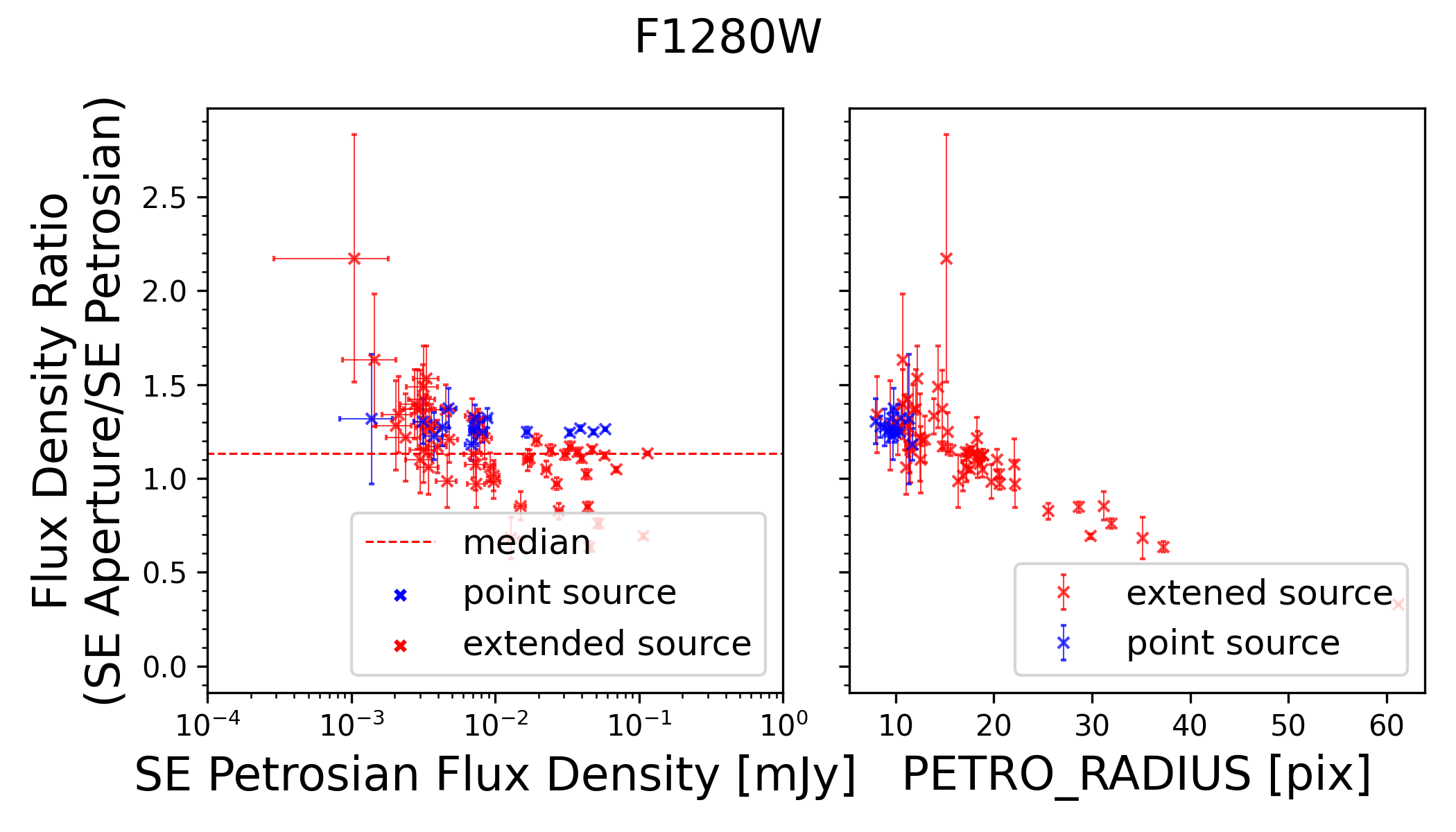}
    \caption{{Same as figure \ref{770ff} but for 12.8$\mu$m band.}}
    \label{1280ff}
\end{figure}

\begin{figure}
    \centering
    \includegraphics[width=\columnwidth]{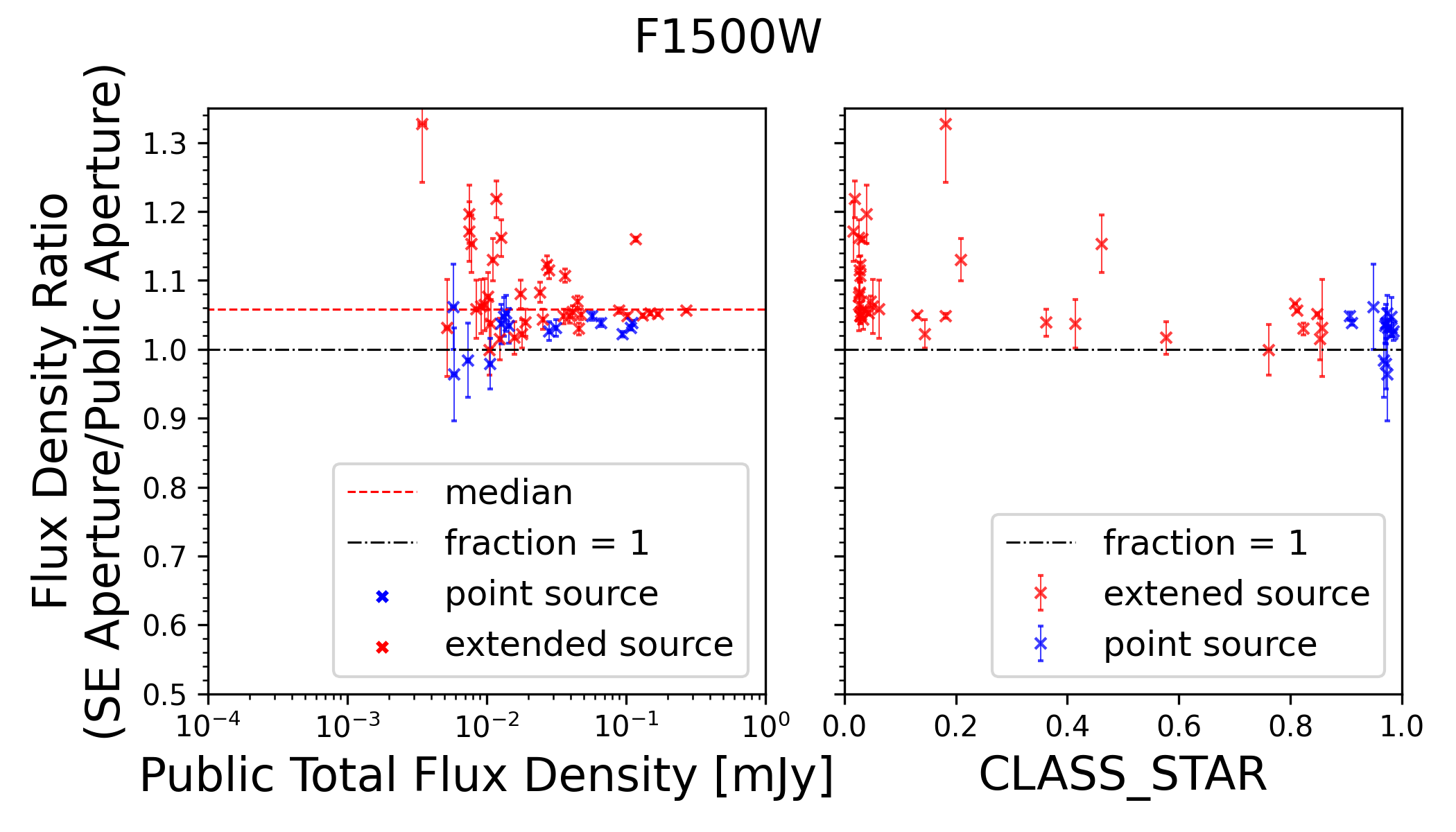}
    \includegraphics[width=\columnwidth]{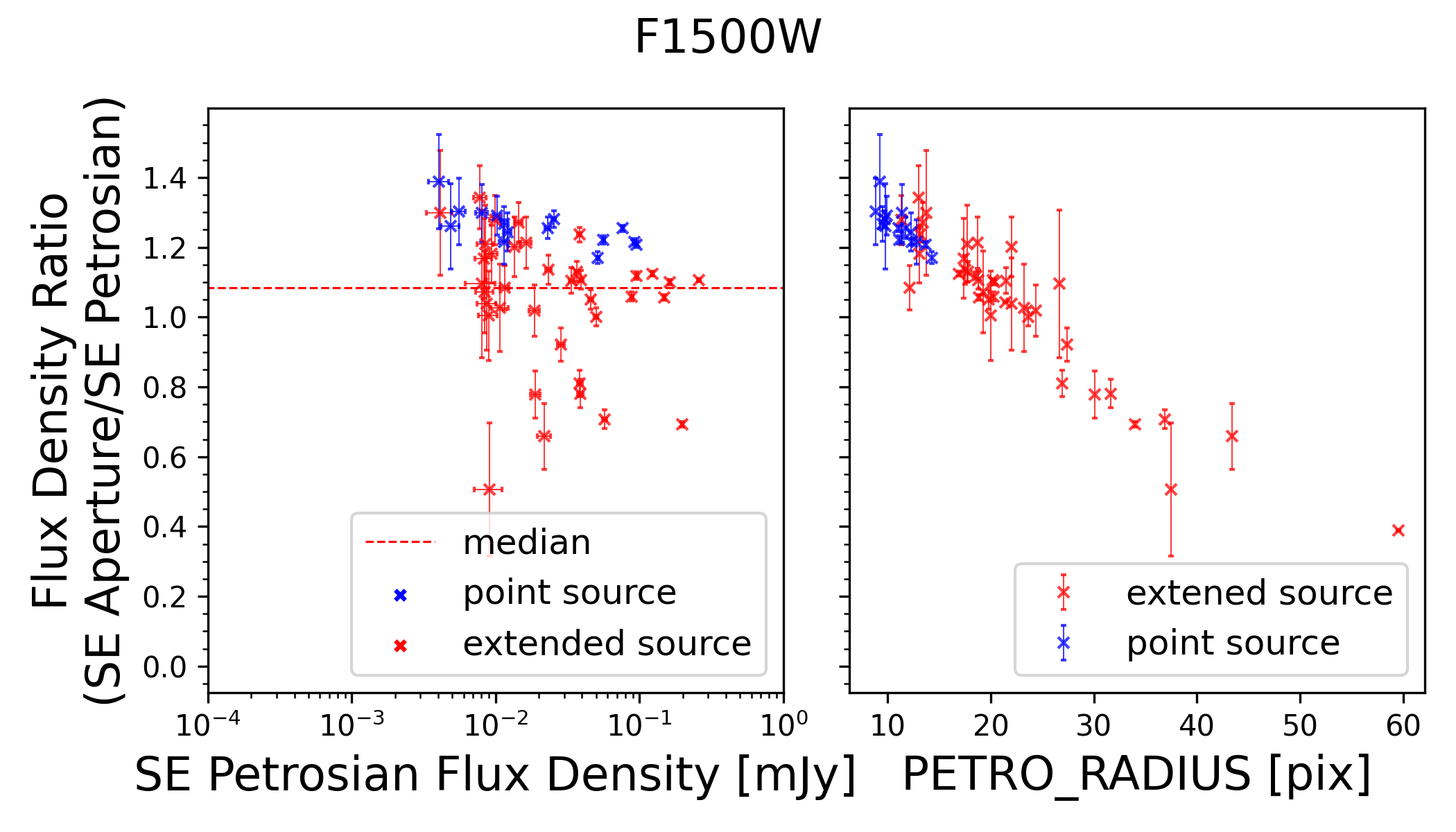}
    \caption{{Same as figure \ref{770ff} but for 15$\mu$m band.}}
    \label{1500ff}
\end{figure}

\begin{figure}
    \centering
    \includegraphics[width=\columnwidth]{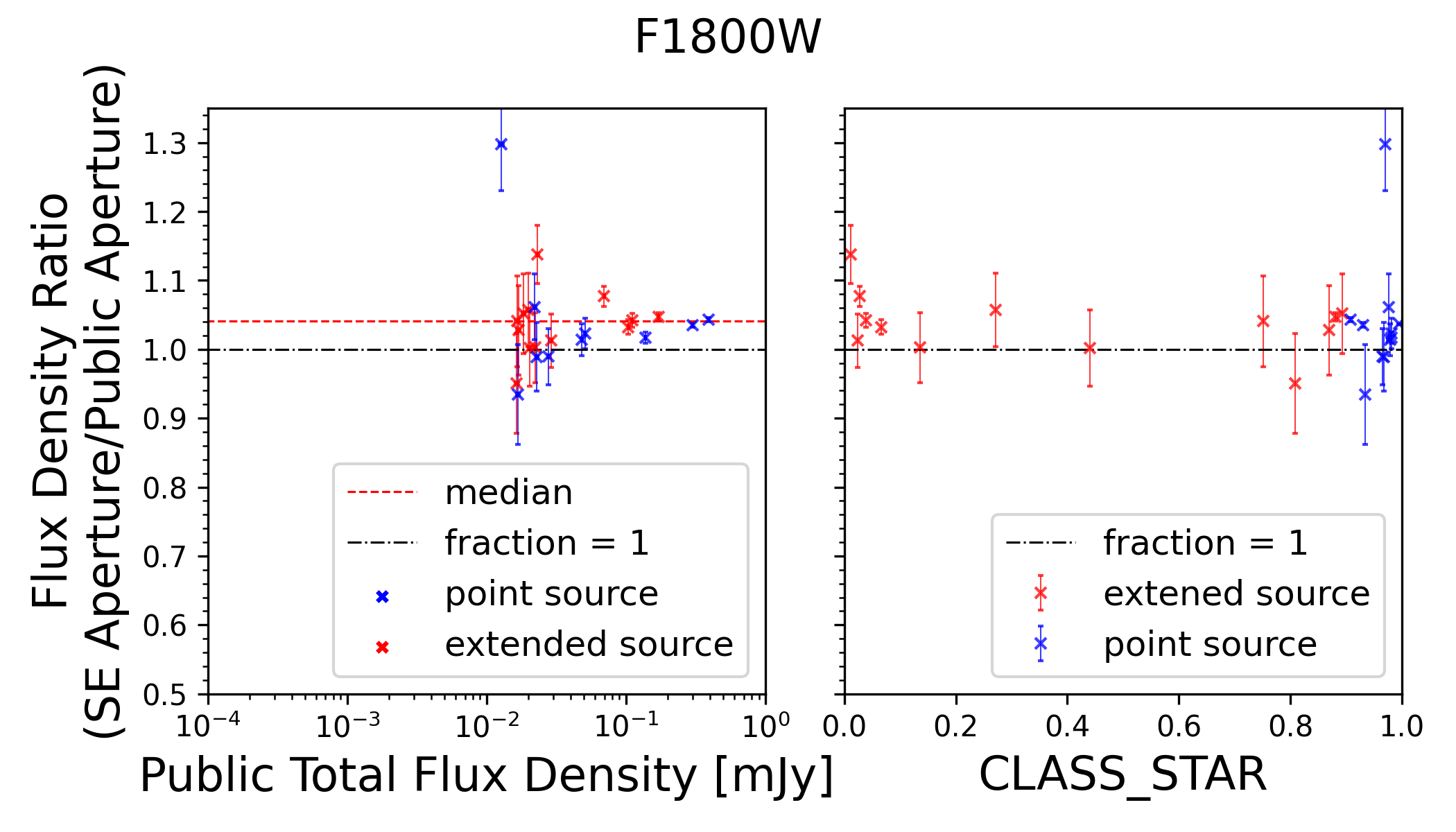}
    \includegraphics[width=\columnwidth]{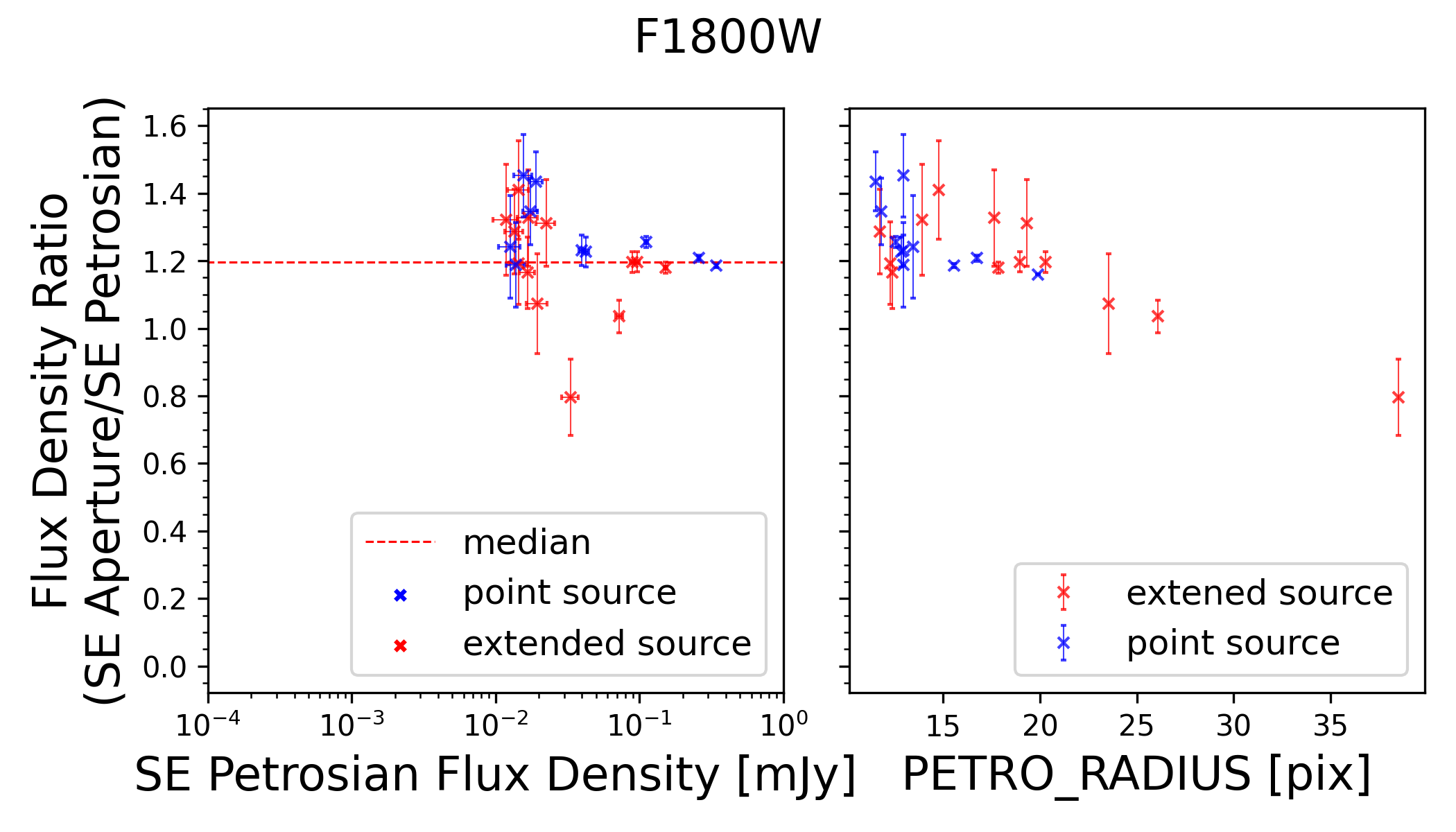}
    \caption{{Same as figure \ref{770ff} but for 18$\mu$m band.}}
    \label{1800ff}
\end{figure}

\begin{figure}
    \centering
    \includegraphics[width=\columnwidth]{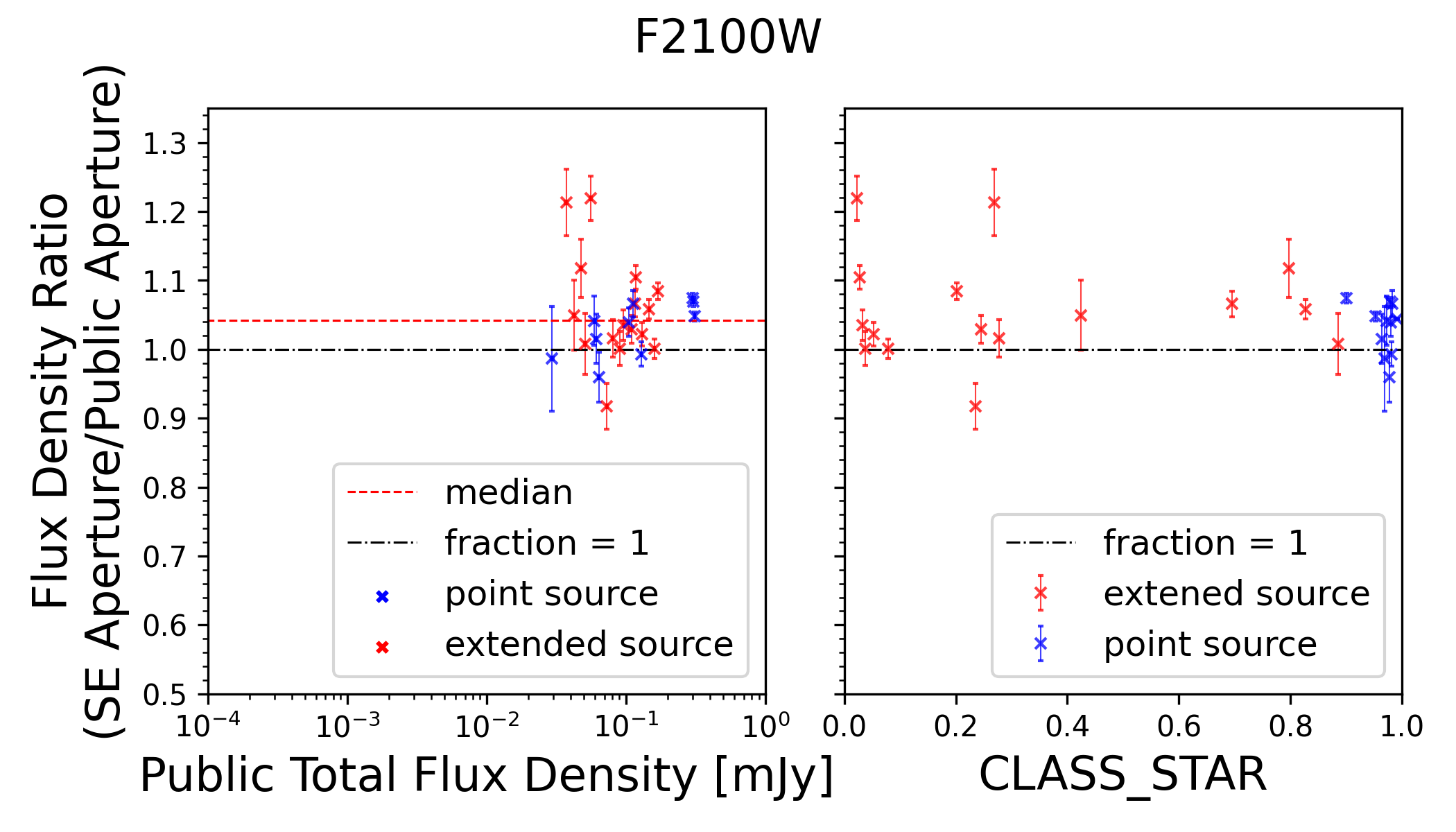}
    \includegraphics[width=\columnwidth]{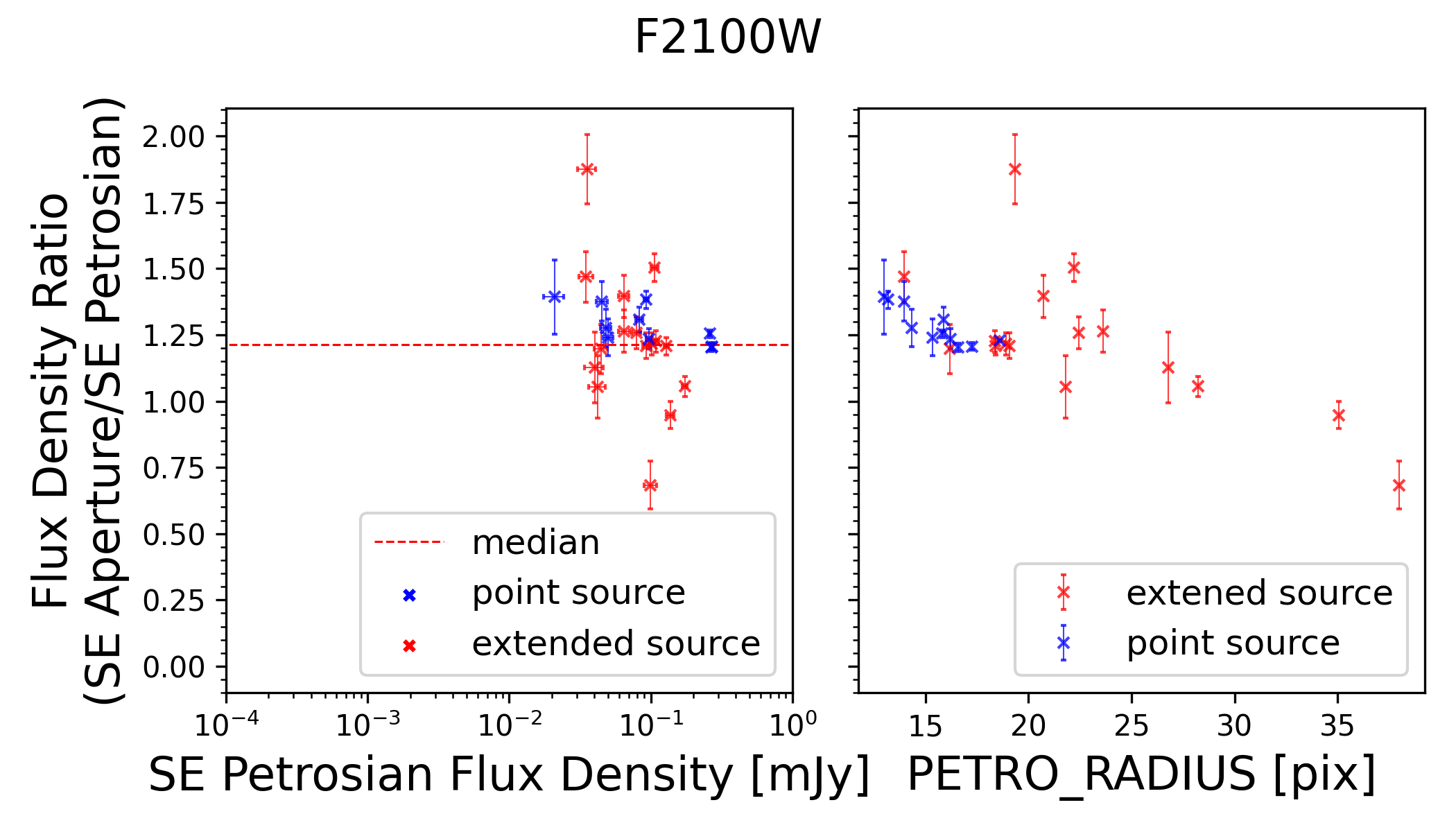}
    \caption{{Same as figure \ref{770ff} but for 21$\mu$m band.}}
    \label{2100ff}
\end{figure}

To illustrate the effectiveness of performing photometry with \textsc{SE}, we compared the aperture-corrected flux measurements obtained from the 70\% enclosed energy apertures, which is used for the public catalog, with the total flux values provided in the public catalog (Fig. \ref{770ff}-\ref{2100ff}). The aperture sizes are documented in the JWST Calibration Reference Data System (CRDS \footnote[2]{\url{https://jwst-crds.stsci.edu/}}). The flux ratio between our photometry and the public catalog is shown in the upper-left panel of Fig. \ref{770ff}-\ref{2100ff}. The resulting flux measurements {have} a median (horizontal dashed lines) that agrees well within 1.3\%, 4.5\%, 4.2\%, 5.8\%, 4.0\%, 4.2\% flux excess/loss in F770W, F1000W, F1280W, F1500W, F1800W, F2100W filters, respectively. The ratio begins to diverge as the fluxes approach the detection limit, that is, the 1-sigma deviation is smaller for the brighter sources. In the upper-right panel, we plot the flux ratio as a function of the CLASS$\_$STAR parameter derived by \textsc{SE}, where the extended sources are selected by CLASS$\_$STAR $\leq$ 0.9, and the remaining sources are considered as point sources. With 70\% enclosed energy aperture photometry, the majority of the samples show no clear deviation from the public catalog for either extended or point sources.

Nevertheless, the fixed aperture size may be too small to recover all the {flux} for some spatially extended sources, since both \textsc{SE} and \textsc{Photoutils} use the same fixed aperture size. To investigate this further, in the lower left panel we compare the fixed aperture flux with the Petrosian flux, also measured by \textsc{SE}, which uses an adaptive aperture. The lower value on the y-axis indicates that the fixed aperture photometry underestimates the fluxes. In addition, the bottom-right panel shows the ratio between the Petrosian radius and the flux ratio between the two photometry methods, as the Petrosian aperture includes more flux for extended sources. The flux ratio between the fixed aperture and Petrosian decreases with increasing source size (PETRO$\_$RADIUS), also indicating that the aperture size is inadequate for extended sources. To avoid the {loss of flux from} extended sources, we decided to use the Petrosian aperture measured with \textsc{SE} for further analysis throughout this paper.

We summarise the parameters specified in \textsc{SE} and \textsc{Photutils} in Table \ref{table:1}.
The parameters in Table \ref{table:1} have been optimized for the source detection of the faintest sources in all six MIRI bands.
Other parameters (DETECT\_MAX/MINAREA, etc.) not listed in Table \ref{table:1} remain the default value of \textsc{SE}.
We also show examples of extracted sources with \textsc{SE} in each filter in the image o001 in the appendix \ref{Appendix}.

\begin{table*}
    \centering
    \begin{tabular}{cp{0.65\columnwidth}cp{0.55\columnwidth}}
        \hline
        Parameter Name && Common to the all filters &\\
        \hline
        DEBLEND$\_$NTHRESH && 48 &\\
        DEBLEND$\_$MINCONT && 0.0008 &\\
    \end{tabular}

    \begin{tabular}{ccccccc}
        \hline
         & F770W & F1000W & F1280W & F1500W & F1800W & F2100W\\
        \hline SN
        DETECT$\_$THRESH [mag/arcsec$^2$] & 1.5 & 1.5 & 1.5 & 1.3 & 1.2 & 1.0 \\
        FILTER$\_$NAME & gauss$\_$2.5$\_$5$\times$5.conv && gauss$\_$3.0$\_$7$\times$7.conv && gauss$\_$4.0$\_$7$\times$7.conv & gauss$\_$5.0$\_$9$\times$9.conv\\
        \hline
    \end{tabular}
    \\
    \caption{Parameters specified in \textsc{Source-Extractor}.}
    \label{table:1}
\end{table*}

\subsection{Completeness}\label{completeness}
To accurately measure the source counts, it is important to correct them for completeness of source detection at each band. 
Therefore, we estimated completeness as a function of flux density.
The final source counts have been corrected for the completeness of our source extraction.

\begin{figure}  
    \centering
    \includegraphics[width=0.7\columnwidth]{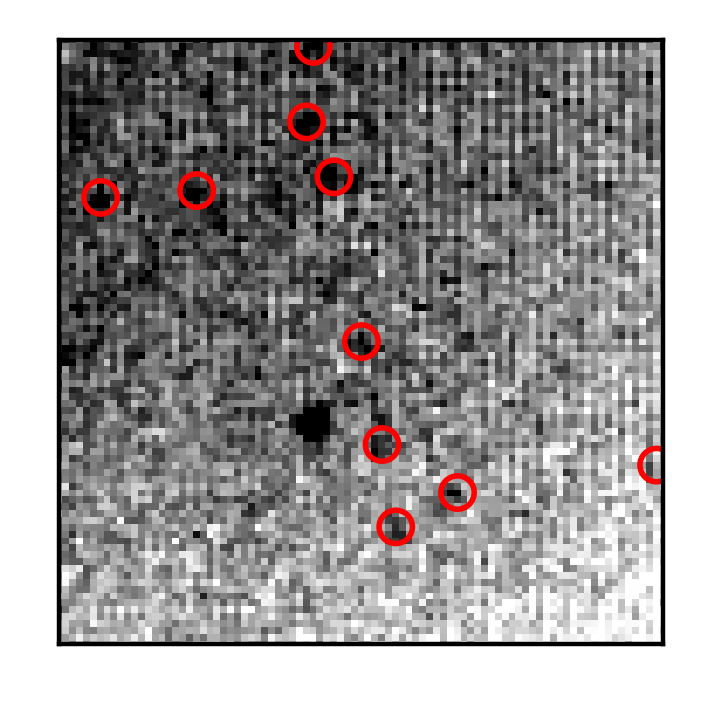}
    \caption{An example of multiple artificial sources inserted from the completeness measurements (indicated by red circles) with fluxes $=1 \:\mu$Jy on the slice of field o001, in the F770W filter.}
    \label{fig.fake}
\end{figure}

We added the artificial sources with a certain range of flux densities {to} the images and examined if they are detected with the same method as we detect the real astronomical sources, as described in \cite{Takagi2012}.
Each image consists of 20 randomly distributed artificial point sources generated from simulated \textit{JWST} MIRI PSF.
All the sources are placed avoiding the edges of the image.
Fig. \ref{fig.fake} shows the example of our artificial sources inserted into the image.
After inserting, we re-run \textsc{SE} with the same set of parameters as the source detection and measure the percentage of artificial sources recovered from the image.
This process is repeated {500} times with varying {the flux range from 0.01 $\mu$Jy to 100 $\mu$Jy} by an increment of 0.1 dex.
In each image, in total, {500,000} artificial sources were examined.
We here assume a flat number distribution for each filter.

\begin{figure}
    \includegraphics[width=\columnwidth]{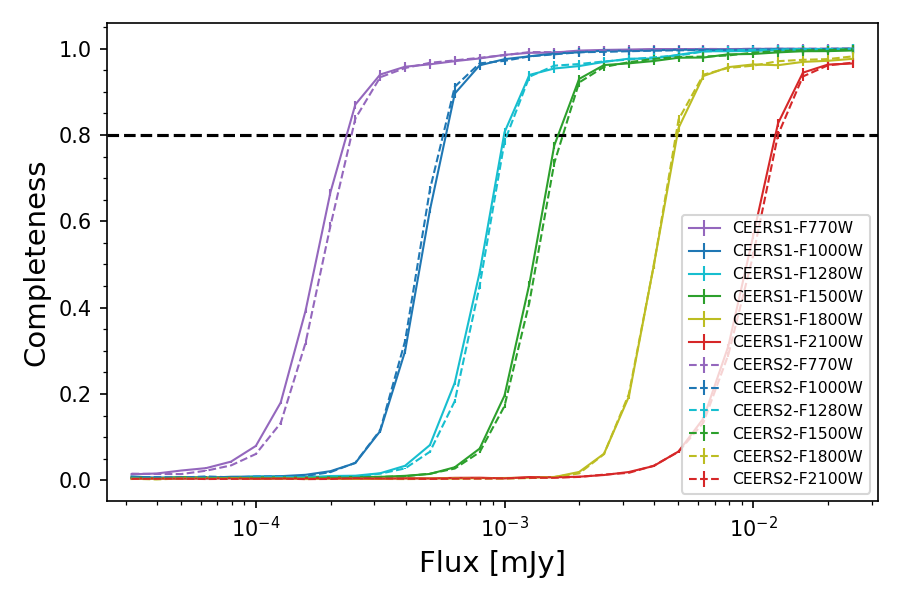}
    \caption{
    The measured completeness of the CEERS field images with a bin width of $\Delta\log (f_{\nu}/{\rm Jy})=0.1$ dex. The results in different filters are plotted in violet (F770W), blue (F1000W), cyan (F1280W), green (F1500W) yellow (F1800W), and red (F2100W). {The solid and dashed lines represent the recovery rate from the o001 and o002 {fields}, respectively.} The black dashed line shows our criterion of 80\% completeness.
    }
    \label{fig:complete}
\end{figure}

In Figure \ref{fig:complete}, we show the derived completeness for CEERS o001 and o002 fields.
The completenesses of the two fields  are matched well across all six filters.

In the o001 (o002) fields, the 80 percent completeness limit reaches {0.25 (0.25), 0.63 (0.63), 1.26 (1.26), 2.0 (2.0), 5.0 (5.0), and 13 (16) $\mu$Jy} in F770W, F1000W, F1280W, F1500W, F1800W, and F2100W filters, respectively.

For comparison, we have scaled the sensitivity of the point source detection limits (SNR=10) from the MIRI instrument Handbook\footnote[3]{\url{http://web.physics.ucsb.edu/~cmartin/data/4clm/MIRI._Cycle1.pdf}} to the effective exposure time of both fields.
The exposure time for each wavelength is 1648.4, 1673.3, 1673.3, 1673.3, 1698.3, and 4811.9 seconds, in F770W, F1000W, F1280W, F1500W, F1800W, and F2100W filters, respectively, corresponding to the flux limits for SNR=10 of 0.24, 0.49, 0.85, 1.4, 3.0, and 5.4 $\mu$Jy. The 80\% completeness limits we obtained are comparable with these numbers. 

\begin{figure}
    \centering
     \includegraphics[width=\columnwidth]{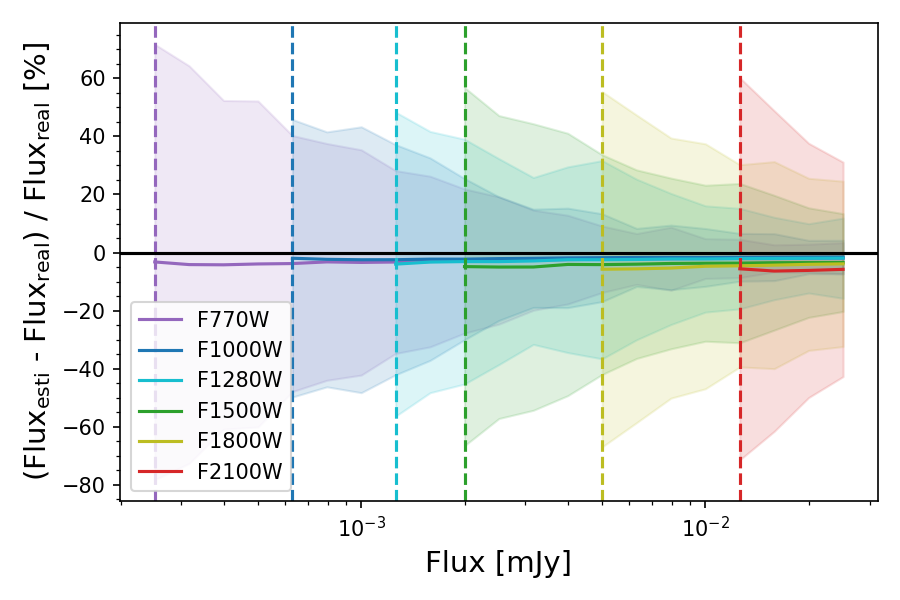}
    \caption{{The difference between the recovered flux from \textsc{SE} and the input value from the simulation. Colour regions show the 1-sigma errors. Colour dashed lines indicate the 80\% completeness limit, above which we use sources for the analyses.}}
    \label{fig:fluxerr}
\end{figure}

During the completeness estimation, we also calculated the accuracy of the flux recovery from \textsc{SE}. As shown in Figure \ref{fig:fluxerr}, the median of flux difference between 'real' (fluxes of injected artificial sources) and 'esti' (fluxes obtained by \textsc{SE}) approaches 0\% after the 80\% completeness limit in every filter. We note, however, there is still a systematic underestimate of $3\pm0.8\%, 2\pm0.6\%, 3\pm0.5\%, 4\pm0.5\%, 5\pm0.4\%, 4\pm0.4\%$ in F770W, F1000W, F1280W, F1500W, F1800W, and F2100W filters, respectively.
The source count of each flux bin is corrected for incompleteness based on our completeness estimation. We only give the corrected source count in the result for sources brighter than 80 percent of completeness.

\subsection{{Reliability}}\label{reliability}
\begin{figure}
    \centering
     \includegraphics[width=\columnwidth]{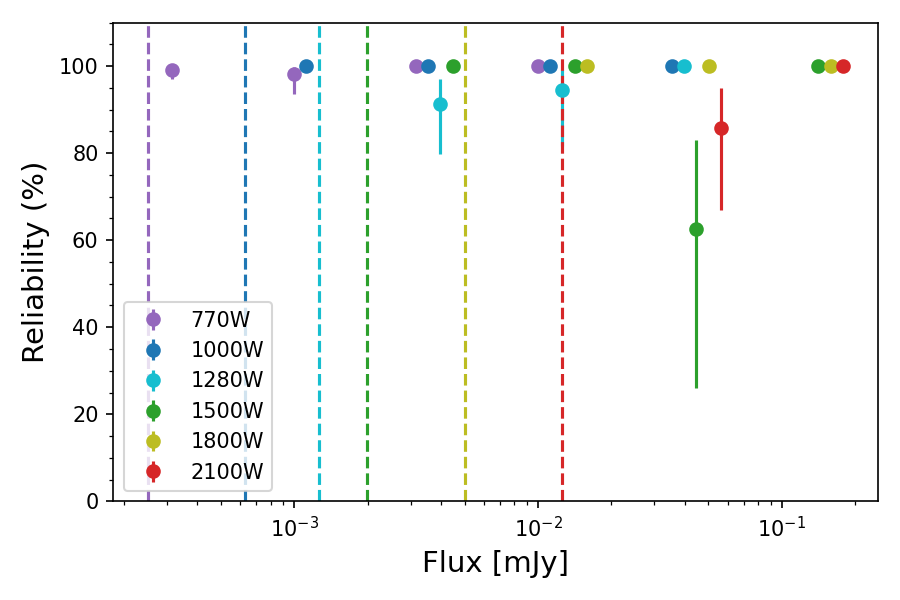}
    \caption{{The reliability as the function of flux with a bin size of 0.5 dex. The reliability is estimated by taking the differences between the sources detected in the original images and the sources detected in the negative images, then dividing the difference by the number of sources detected in the original images. To prevent overlapping, dots in the different filters are each offset by 0.05 dex. Colour dashed lines indicate the 80\% completeness limit for each filter.}}
    \label{fig:reliable}
\end{figure}

To confirm the credibility of the extracted sources, we perform the same source extraction procedure on negative images as on the original positive images. The negative images are created by multiplying the {signal} in the original JWST MIRI images by $-1$. We have removed the artificial spikes near the brightest star in {the} field 'o001' and the spiral galaxy in the lower right corner in the field 'o002'. The reliability is estimated by taking the differences between the sources detected in the original images and the sources detected in the negative images, then dividing the difference by the number of sources detected in the original images, $(N_{pos}-N_{neg})/N_{pos}$. We have combined the number of sources detected in the 'o001' and 'o002' fields for a better evaluation of the overall reliability. Figure \ref{fig:reliable} shows the reliability as a function of the flux above 80\% completeness in each filter. For most filters, we find that the reliability of our sources is above 95\%. The dip at 15 $\mu$m is mainly due to the binning, where fewer sources are obtained in the positive image at this flux range. The 21 $\mu$m image generally has the worst reliability of the filters due to the lower resolution. Table \ref{table:2} summarises the average reliability of each filter.

\begin{table}
    \centering
    \begin{tabular}{cc}
        \hline
        Filter & Reliability \\
        \hline 
        F770W & 99\% \\
        F1000W & 100\% \\
        F1280W & 95\% \\
        F1500W & 95\% \\
        F1800W & 100\% \\
        F2100W & 72\% \\
        \hline
    \end{tabular}
    \caption{The average reliability {above} the 80\% completeness limit in each filter.}
    \label{table:2}
\end{table}

\section{Results and discussion}\label{S:res}
\begin{figure}
  \begin{center}
  \includegraphics[width=\columnwidth]{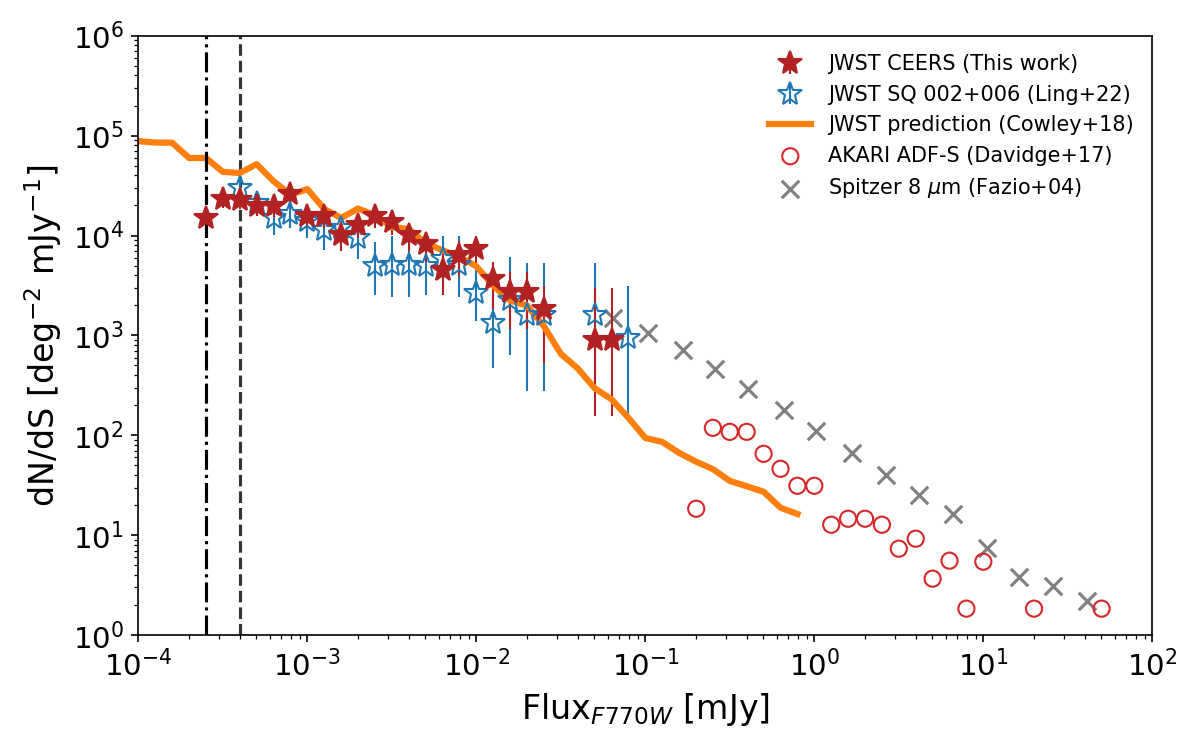}
  \includegraphics[width=\columnwidth]{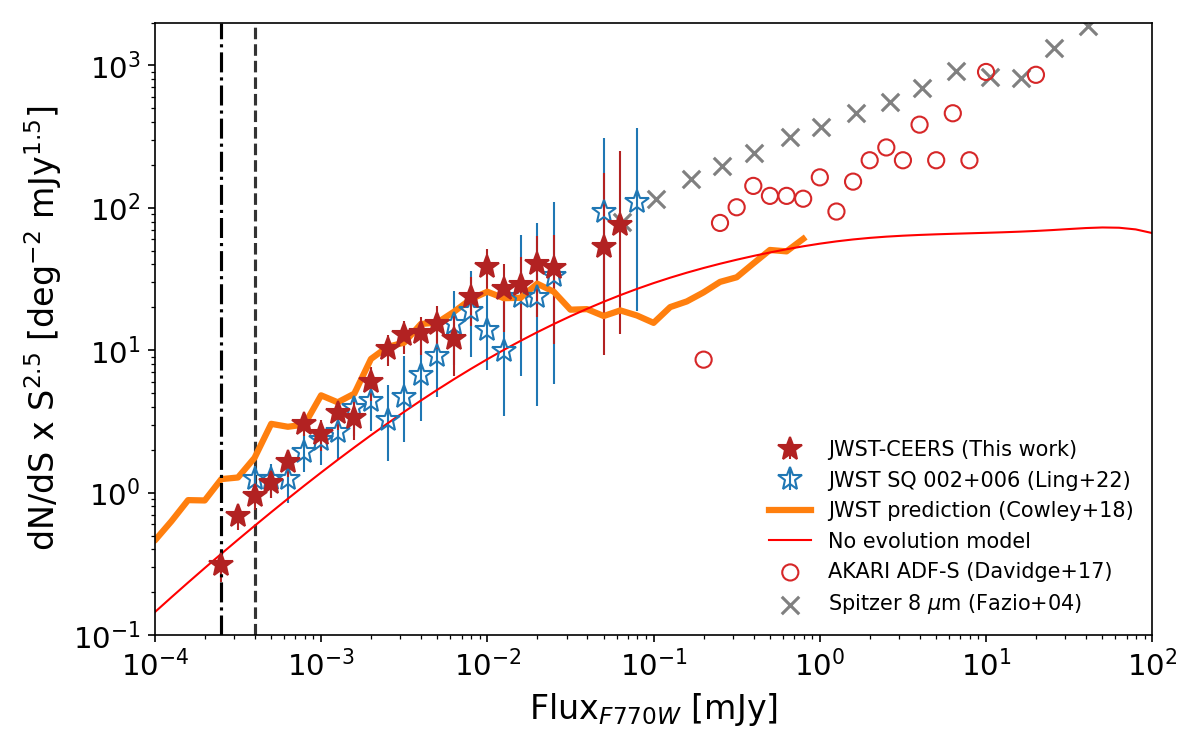}
    \caption{Source counts in the 7.7-$\mu$m band (F770W) respect to \citep[][]{Ling2022}.
    The top panel shows the number of sources per deg$^{2}$ in each flux bin of $0.1$ dex ($\Delta\log (S_{\nu})=0.1$).  
      The bottom panel shows the differential source counts normalized to the Euclidean space.
    Red solid stars indicate the source counts using the \textit{JWST CEERS} data in this work.
    {Blue open stars present the source counts from Stephen's Quintet field with \textit{JWST} \citep[][]{Ling2022}.}
    The orange line shows the model prediction with \textit{JWST} from \citet{Cowley2018}. 
    The \textit{Spitzer} 8 $\mu$m source counts for the Bo\"{o}tes field ($\times$ symbol) from \citet{Fazio2004} is compared as well, where stars dominate the bright end, {making the source count level off from Cowley's model.}
    The black vertical dot-dashed line indicates the \textit{JWST} 80$\%$ completeness limit in the deepest field in this work.
    Similarly, the grey vertical dashed line indicates the \textit{JWST} 80$\%$ completeness level of Stephen's Quintet from \citet[][]{Ling2022} for comparison. 
   {The red solid line indicates the no-evolution model inferred from \citet{Gruppioni2011}.}
    } 
    \label{SC_07}
   \end{center}    
\end{figure}

\begin{figure}
  \begin{center}
  \includegraphics[width=\columnwidth]{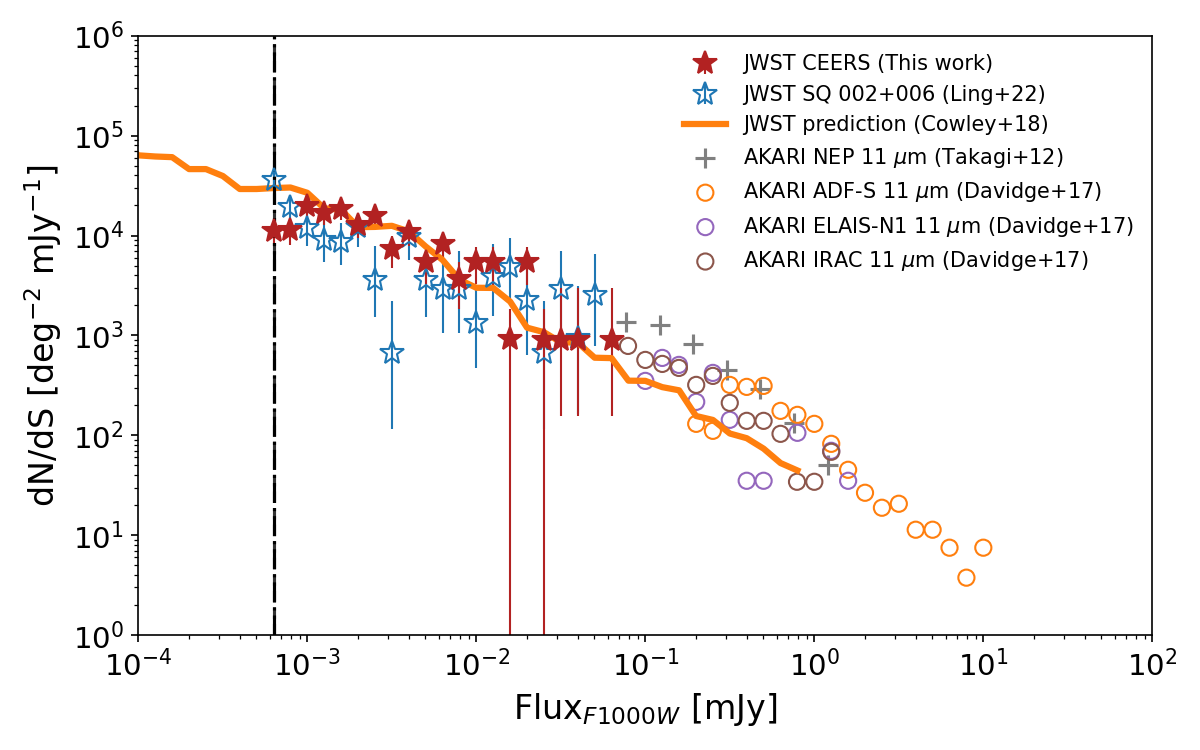}
  \includegraphics[width=\columnwidth]{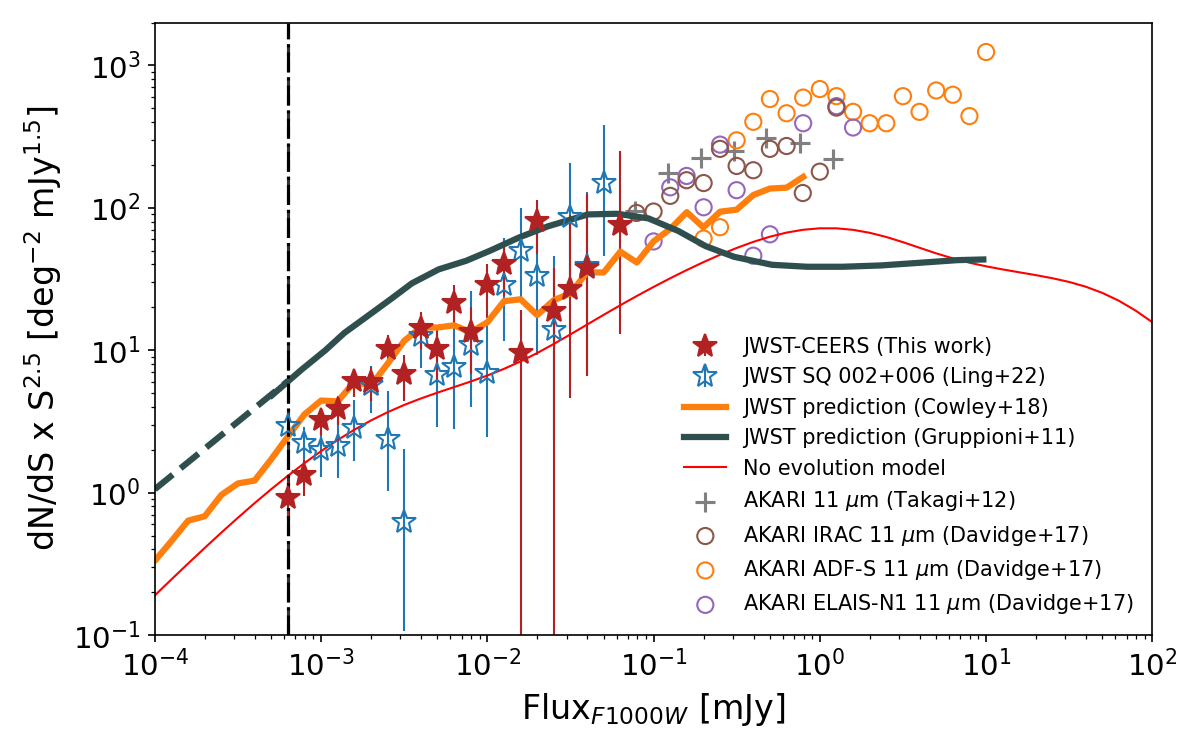}
    \caption{Same as Figure \ref{SC_07}, but for 10-$\mu$m band (F1000W).
    The dark green curve is the model prediction from \citet{Gruppioni2011}, linearly extrapolated to fainter flux in the dark green dashed line. 
    The S11 band (11$\mu$m) of \textit{AKARI} source counts ($+$ symbol) from \citet{Takagi2012} is also presented for comparison.
    }
    \label{SC_10}
 \end{center}    
\end{figure}

\begin{figure}
  \begin{center}
  \includegraphics[width=\columnwidth]{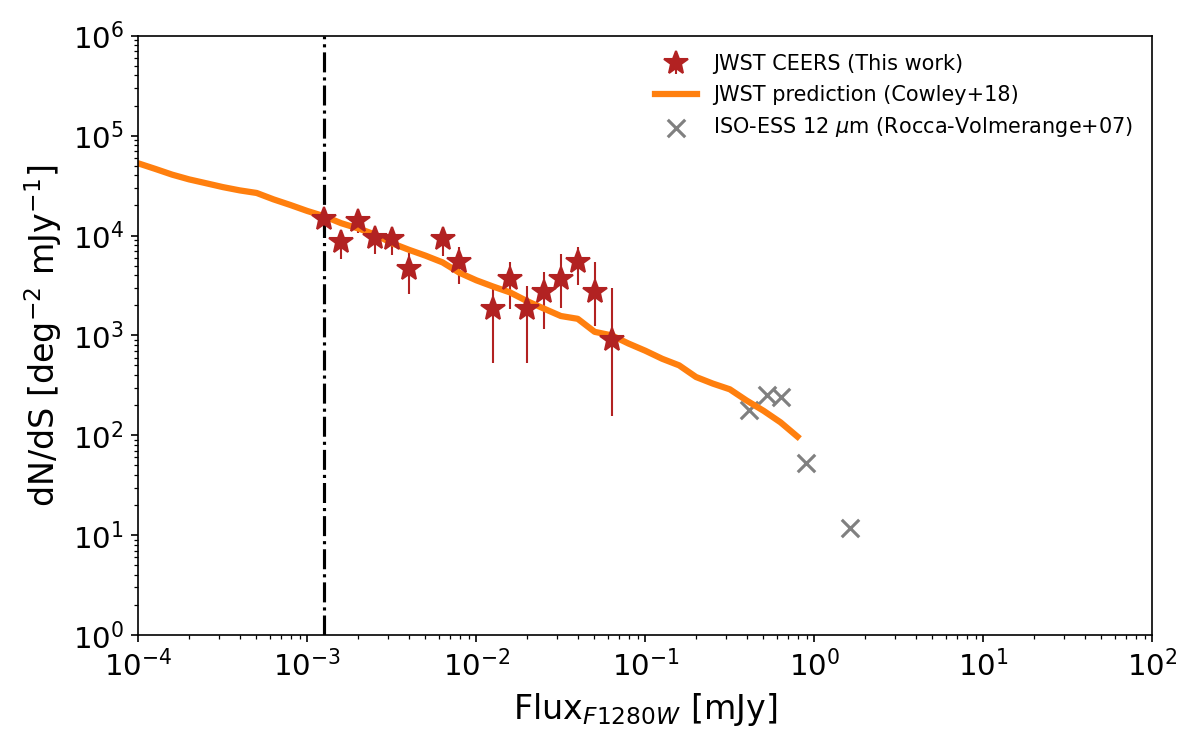}
  \includegraphics[width=\columnwidth]{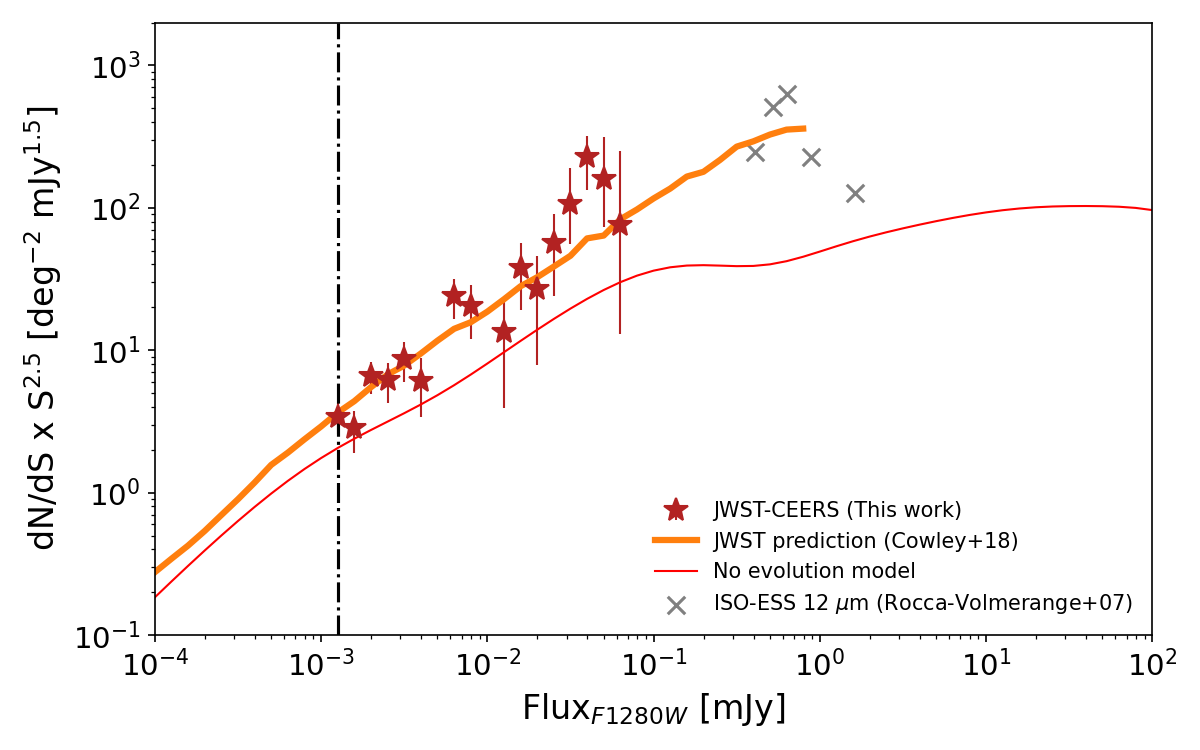}
    \caption{Same as Figure \ref{SC_07}, but for 12.8-$\mu$m band (F1280W). 
    Red open stars indicate the source counts for \textit{JWST CEERS} data from this work only.
    We over-plot the 12$\mu$m result from ISO-ESS survey performed by ESA's \textit{Infrared Space Observatory} (ISO) \citep{Rocca2007}.
    } 
    \label{SC_12}
   \end{center}    
\end{figure}

\begin{figure}
  \begin{center}
  \includegraphics[width=\columnwidth]{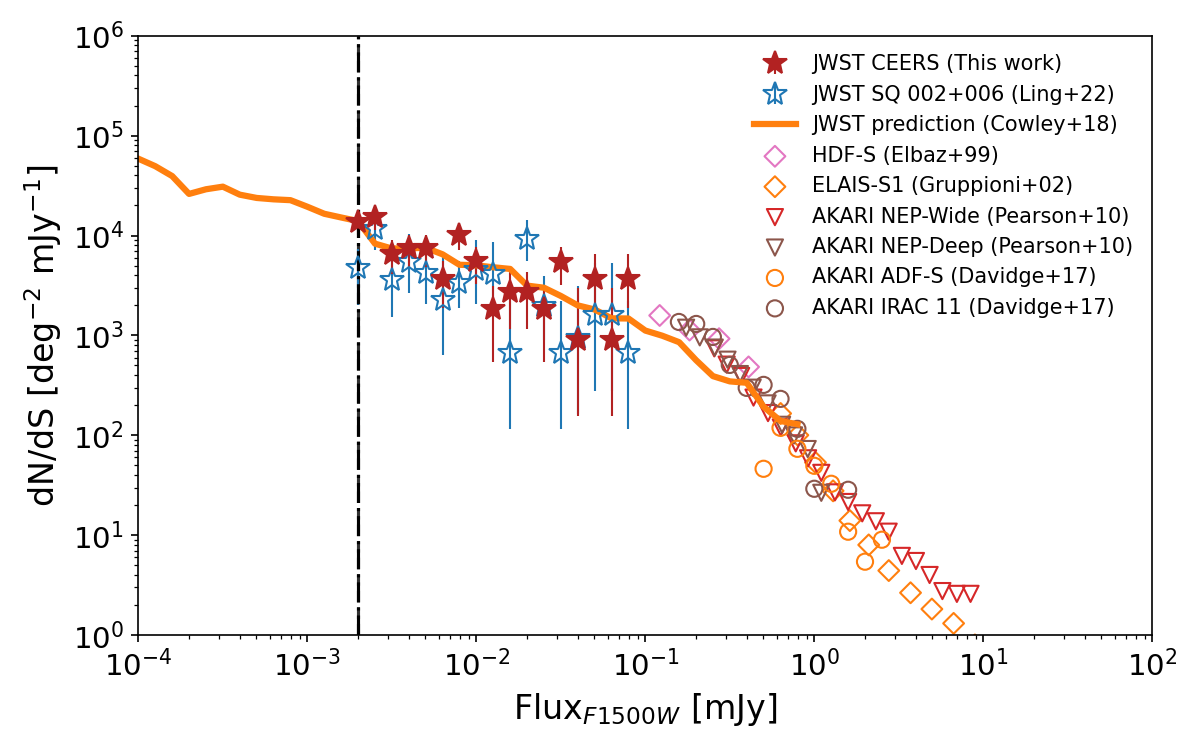}
  \includegraphics[width=\columnwidth]{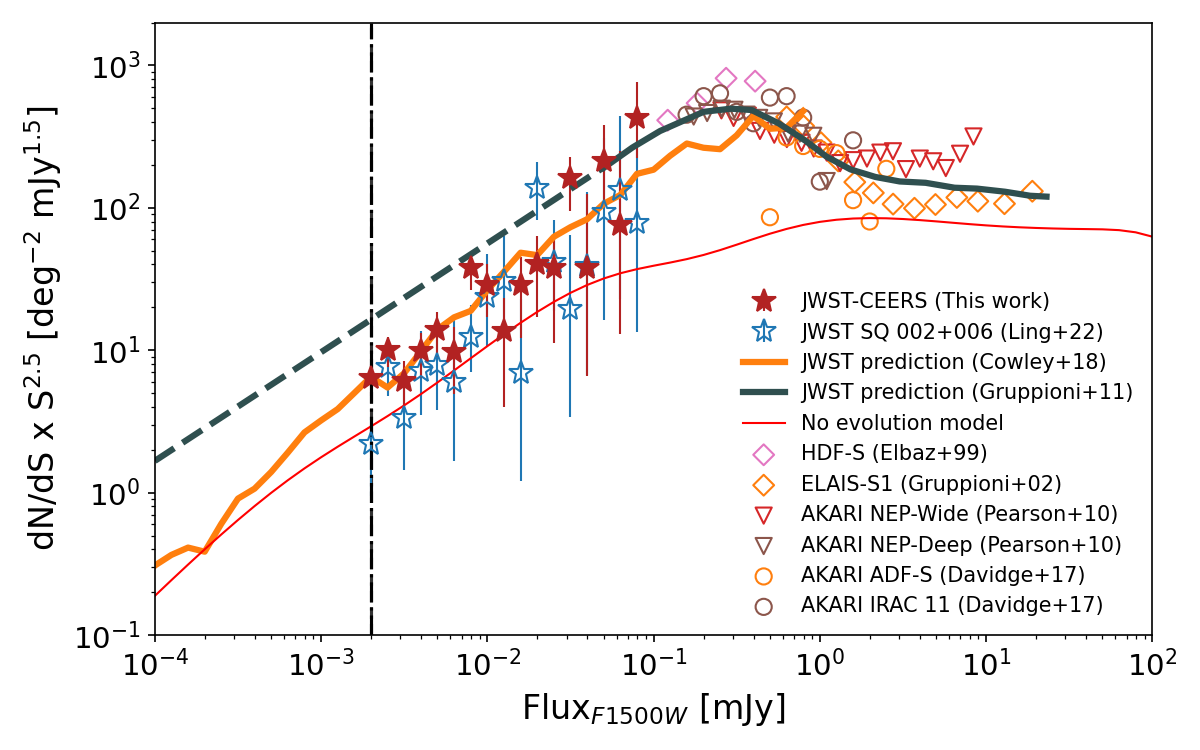}
    \caption{{Same as Figure \ref{SC_07}, but for 15-$\mu$m band (F1500W). We show \citet{Gruppioni2011} prediction and its linear extrapolation at the low-flux end as a dark green solid line and dashed line, respectively. The source counts in precious work are included as: \citet[][pink diamonds]{Elbaz1999}, \citet[][orange diamonds]{Gruppioni2002}, \citet[][inverted triangles]{Pearson2010}.}}
    \label{SC_15}
  \end{center}
\end{figure}

\begin{figure}
  \begin{center}
  \includegraphics[width=\columnwidth]{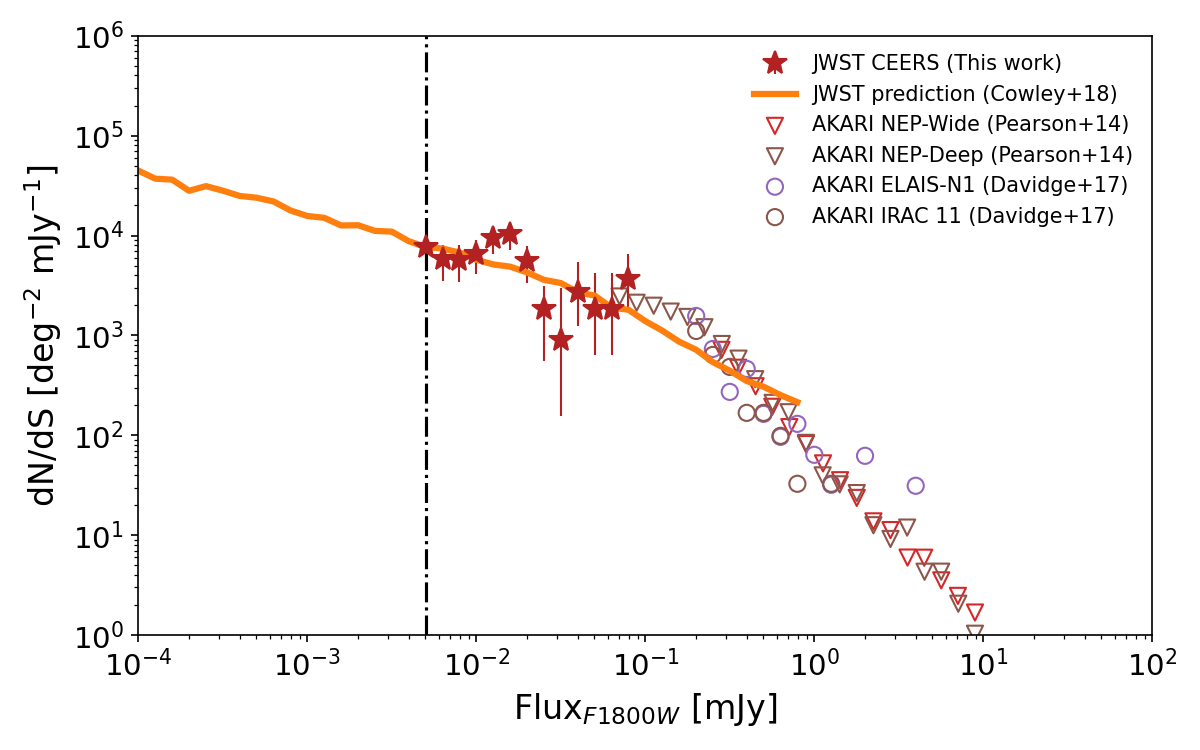}
  \includegraphics[width=\columnwidth]{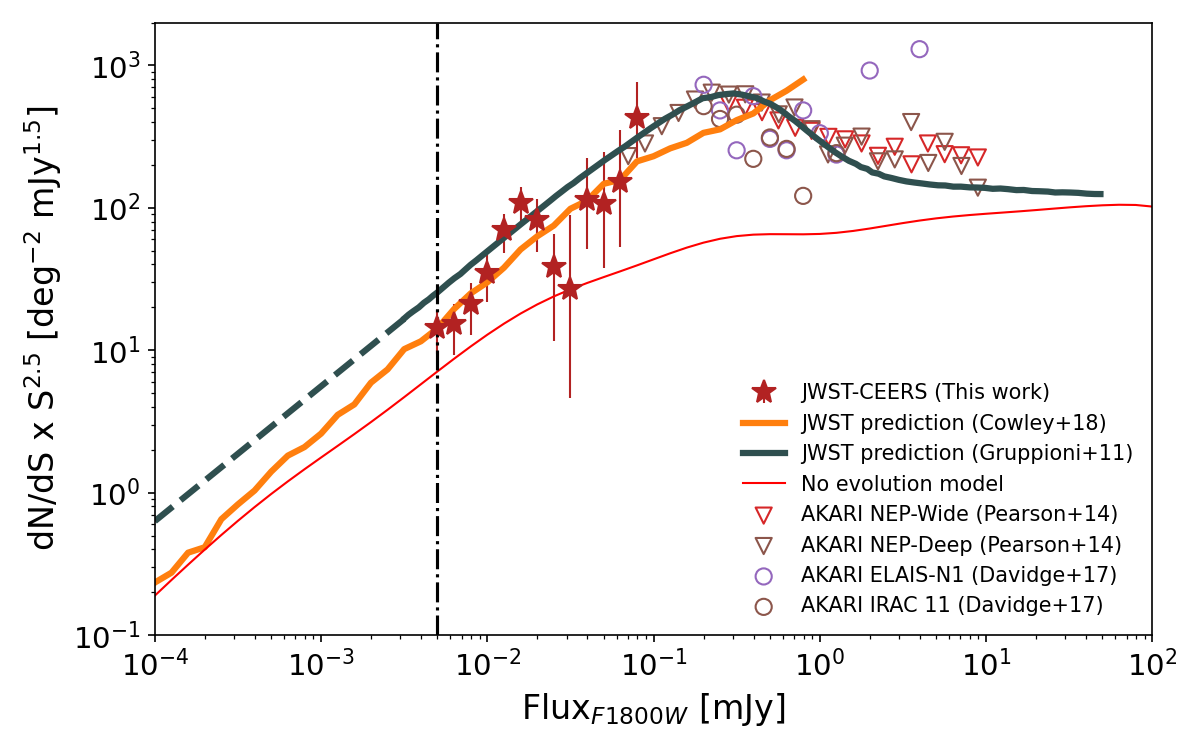}
    \caption{
    Same as Figure \ref{SC_07}, but for 18-$\mu$m band (F1800W).
    Red open stars indicate the source counts for \textit{JWST CEERS} data from this work only.
    We show \citet{Gruppioni2011} prediction and its linear extrapolation at the low-flux end as a dark green solid line and dashed line, respectively.
    The inverted triangles are the data points from \textit{AKARI} NEP survey at 18-$\mu$m \citep{Pearson2014}.
   }
    \label{SC_18}
  \end{center}
\end{figure}

\begin{figure}
  \begin{center}
  \includegraphics[width=\columnwidth]{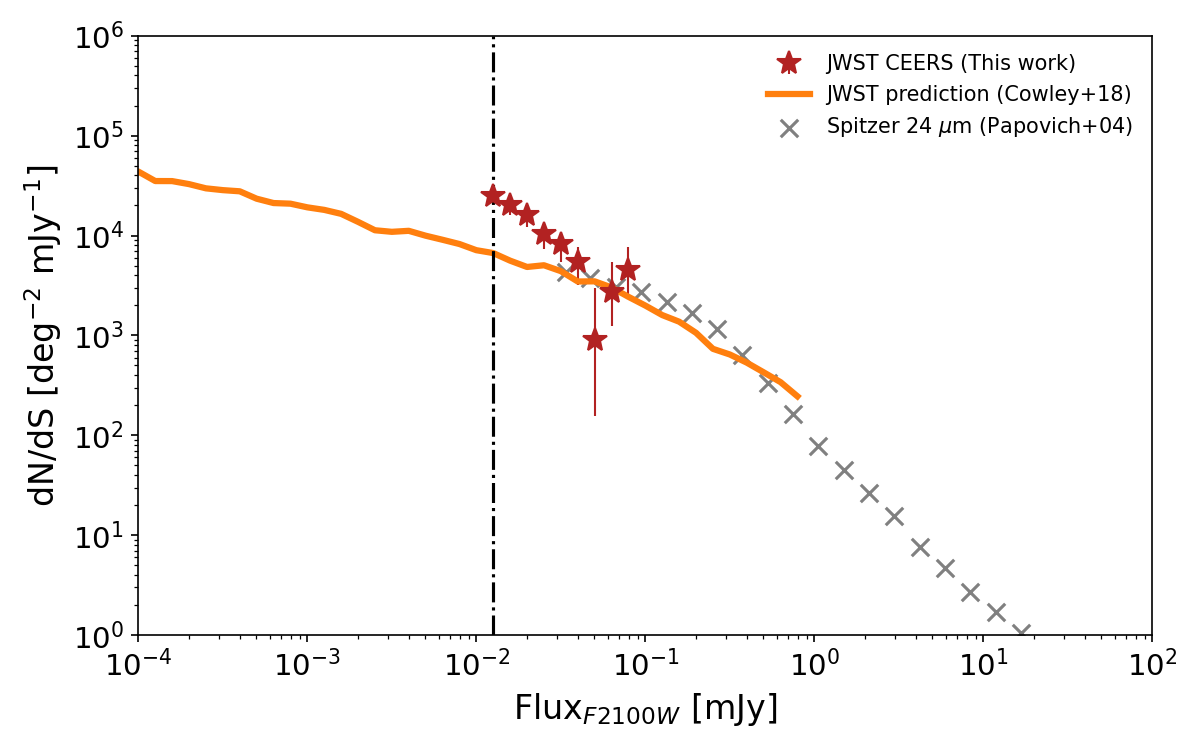}
  \includegraphics[width=\columnwidth]{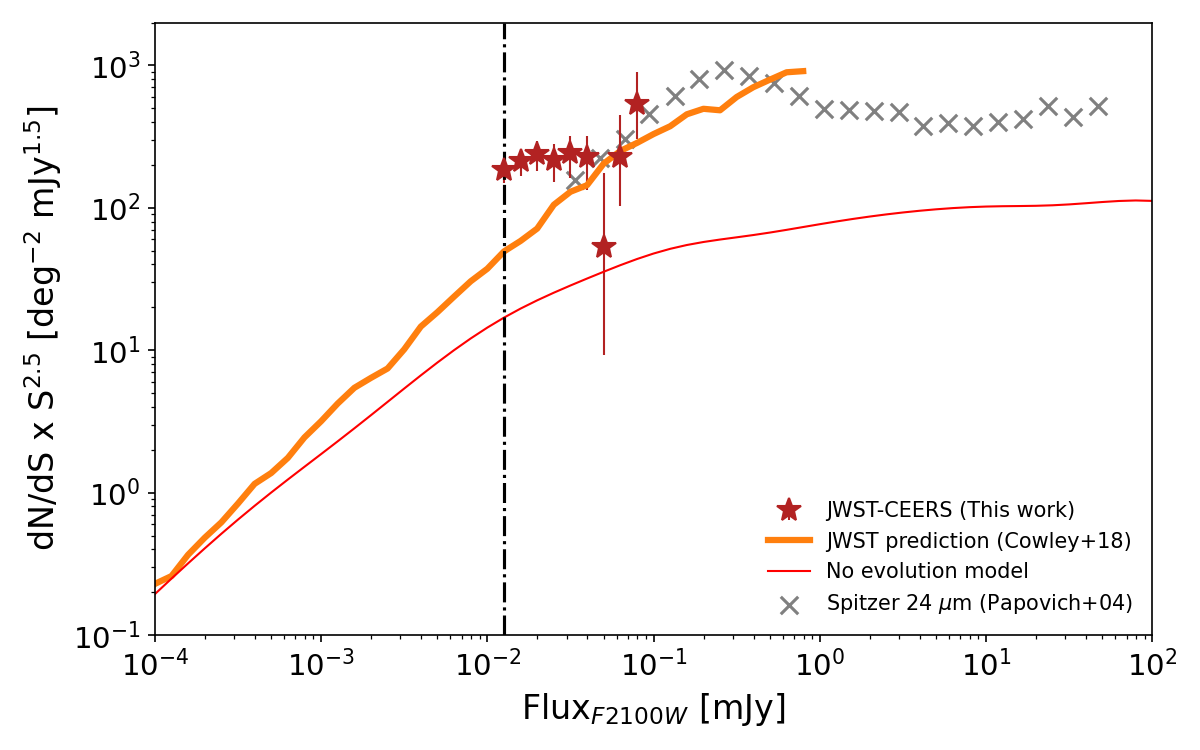}
    \caption{
    Same as Figure \ref{SC_07}, but for 21-$\mu$m band (F2100W).
    Red open stars indicate the source counts for \textit{JWST CEERS} data from this work only.
    The grey crosses are the observed source counts with \textit{Spitzer} at 24-$\mu$m from \citet{Papovich2004}.
    {Note that we did not apply any flux correction between 24-$\mu$m and 21-$\mu$m.}
   }
    \label{SC_21}
  \end{center}
\end{figure}

We present the {derived} source counts in the CEERS o001 and o002 fields in Figs. \ref{SC_07}-\ref{SC_21}.
For F770W, F1000W, and F1500W, we overlay the source counts with those from Stephan's Quintet fields \citep[][]{Ling2022}.
The total area in Stephan's Quintet fields is 16688 arcsec$^2$ ($\sim$4.64 arcmin$^2$), {which is} larger than that covered in this work.
We derived the Poisson error for our source count estimation at each flux bin as follows.
If the number of extracted sources from each CEERS field within one flux bin is fewer than 6, we adopt the error values provided by \citet{Gehrels1986} for {an improved} estimation.
In the resulting source counts, the non-uniformity at longer wavelengths is due to a smaller number of sources {being detected} in those bands. {This could be attributed} to the binning of the data, {and} the lower resolution and sensitivity at longer wavelengths

\begin{figure}
  \begin{center}
  \includegraphics[width=\columnwidth]{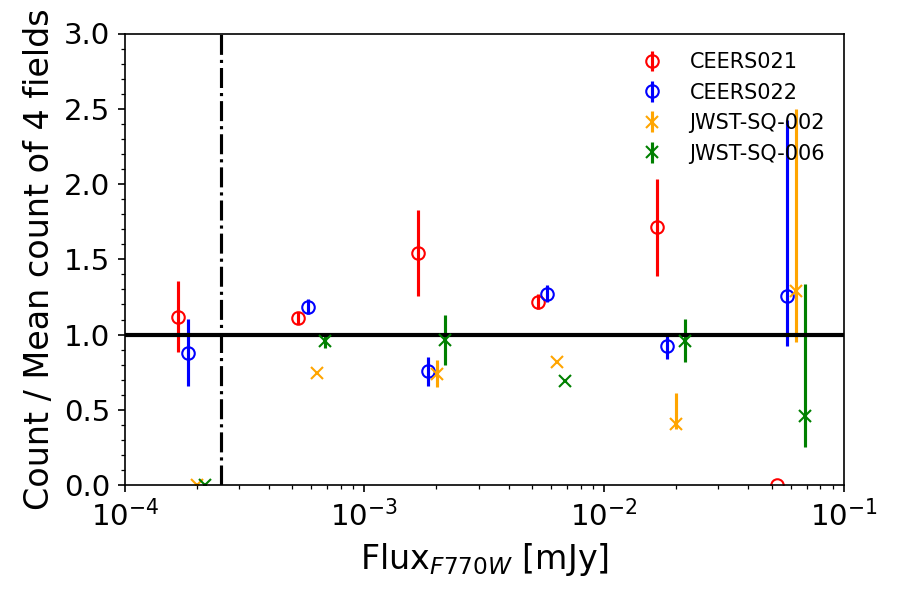}
  \includegraphics[width=\columnwidth]{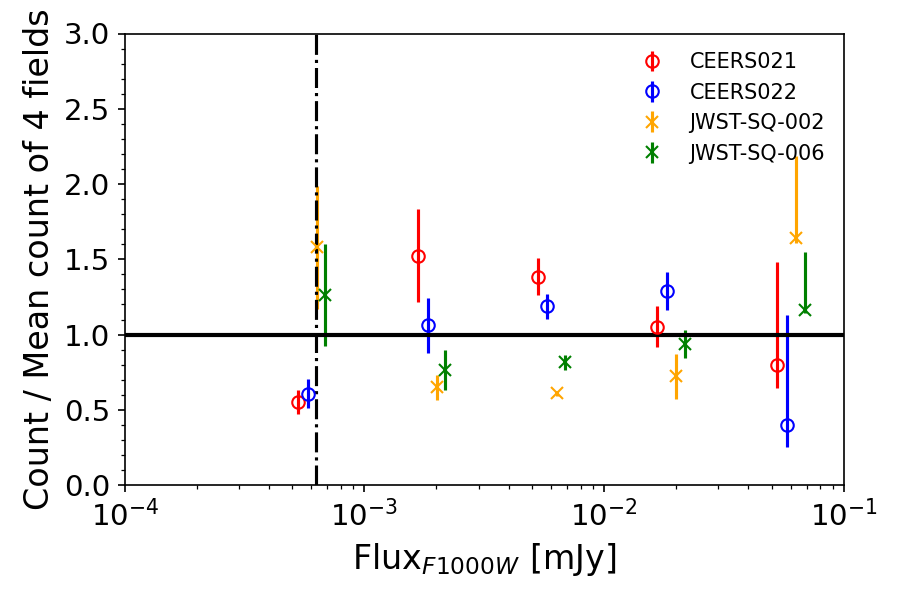}
  \includegraphics[width=\columnwidth]{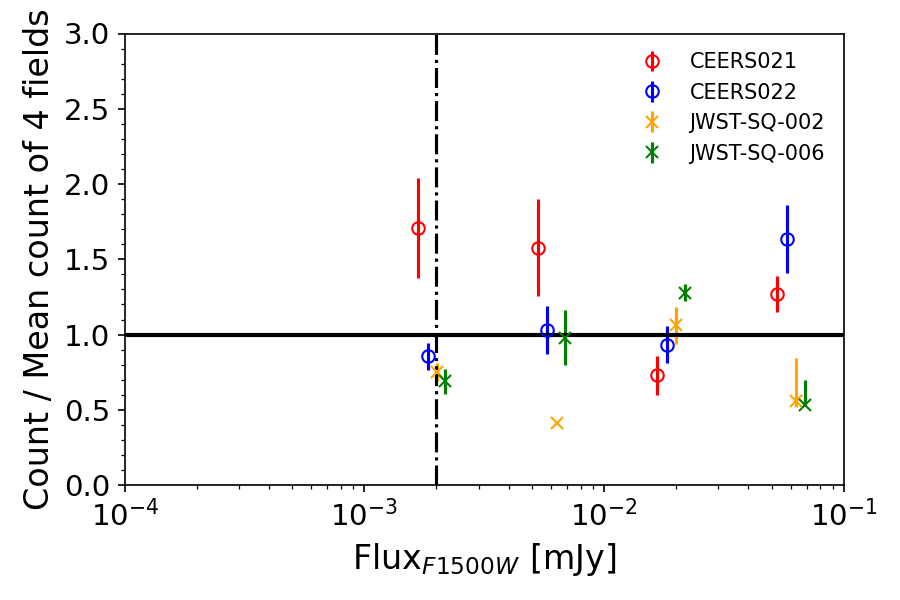}
    \caption{{The field-to-field comparison between two Stephan's Quintet fields from \citet{Ling2022} and two CEERS fields (this work). The data points represent the average of each bin and source counts with the bin size of 0.5 dex in F770W, F1000W, and F1500W. The error of these fluctuations is the Poisson noise of the source counts. The derived field-to-field fluctuation is between 30\% to 180\%.}}
    \label{LSS}
  \end{center}
\end{figure}

Compared with \citet[][]{Ling2022}, the CEERS field is {better suited for the purpose of extra-galactic source counts.} This field has been obtained for the extra-galactic studies, {resulting in fewer contamination from} the foreground objects. With this CEERS field data, we are able to explore deeper than Stephen's Quintet {fields} used by \citet[][]{Ling2022}.
{At} the shortest wavelength 7.7$\mu$m (Fig. \ref{SC_07}), our 80\% completeness level {reaches} $\sim$2 times deeper than \citet{Ling2022}.
Our 80\% completeness level in the other two bands (Fig. \ref{SC_10} and \ref{SC_15}) are both comparable to \citet{Ling2022}'s numbers.

We {note} that the observed source counts are high {than} the no-evolution model (red solid line) shown in Figs \ref{SC_07}-\ref{SC_21}. {One possible explanation is that} the assumption of no evolution in the luminosity function (LF) may not be valid for all galaxy populations. The no-evolution model assumes that the comoving luminosity function remains equal to the local one at all redshifts, but this may not be true if galaxies undergo significant evolution in their properties over time. For example, if star formation rates were higher in the past, we might expect to see more luminous galaxies at high redshifts (Figures 4 and 8 in \citet{Goto2010}). Thus, the observed number of evolutionary models will be higher compared to the no-evolution model. Therefore, the no-evolution model can only be a baseline for comparison with more complex empirical models (which allow for LF evolution).

In all six broad band filters, \textit{JWST} {continues} the work from its predecessors {such as} \textit{Spitzer} \citep{Fazio2004} or \textit{AKARI} \citep{Takagi2012}, and the JWST counts still continue down below the $\sim\mu$Jy levels {at} shorter wavelengths (7.7 and 10 $\mu$m).
At brighter flux of >0.1mJy, {the deviation between models and observed data from \citet{Fazio2004} is due to the contamination from Milky Way stars.}
The newly observed data seems to be well connected with brighter source counts from the previous studies such as \citet{Rocca2007}, \citet{Elbaz1999}, \citet{Gruppioni2002}, \citet{Pearson2010}, \citet{Pearson2014}, \citet{Papovich2004} and \citet{Davidge2017} in all filters.
The prediction from the dark matter simulation combined with the SED models \citep[][orange curves]{Cowley2018} appears in {reasonable agreement} in all the observational measurements.
In Figs. \ref{SC_07}, \ref{SC_15} and \ref{SC_18}, the linear extrapolation of the prediction \citep{Gruppioni2011} based on previous generation satellites \citep[\textit{Spitzer}/\textit{AKARI};][]{Gruppioni2010} are over-plotted in the dark green line, and agree with the \textit{JWST} measurements within the errors.

The only exception is Figure \ref{SC_10}, where we compare the JWST source counts in 10 $\mu$m with the model prediction from \citet{Cowley2018} and \citet{Gruppioni2011}.
The former model combined dark matter simulation (GALFORM) with the galaxy SED models (GRASIL) to obtain the prediction, while the latter relies on the parameters that have been constrained by the observables from pre-JWST surveys in the mid- and far-IR.
In Fig. \ref{SC_10} at the fainter flux of $<0.01$ mJy, the \citet{Gruppioni2011} model over-predicts IR sources by a factor of $\sim3$, while the slope is still consistent.

In Fig. \ref{770ff}, the shape of Cowley's model in differential source counts is not seen in wavelengths other than 7.7 $\mu$m. According to the discussion in \citet{Cowley2018}, the burst component, i.e., the star-burst component becomes important in the wavelength > 10$\mu$m at the high-flux end, due to the dust-attenuated energy from nearby galaxies being re-emitted at observed wavelengths that are larger or equal to 10 $\mu$m. However, the starburst component is less dominant {at} 7.7 $\mu$m. At 7.7 $\mu$m, starburst galaxies have a prominent PAH emission, which is difficult to be accurately modeled. Therefore, one possibility is that the contribution from the starburst component needs to be improved in \citet{Cowley2018} at 7.7 $\mu$m. {This might account for the drop observed in Cowley’s model at 0.1 mJy, a feature that is solely evident in Fig. \ref{770ff}.}

{At} all wavelengths, even if we have consistent results with the model predictions, the parameters in the \citet{Gruppioni2011} model has inevitable degeneracy.
The source counts {themselves do} not provide specific information about any particular type of IR population, {making it difficult} to be connected to the LF of a certain type of galaxies.  
In this sense, the degeneracy in luminosity or density evolution, or the faint-end slopes of LFs exist in the source counts. 
{Deviations seen} in this work may suggest an evolution towards the fainter end, {or the need to modify} evolutionary parameters in the currently available models {in order to} describe all the infrared sources ranging from Jy to sub-$\mu$Jy levels.
In the near future, we will be able to use photometric/spectroscopic redshifts and more advanced SED-fitting techniques with brand-new template libraries to unwrap/improve these parameters or models \cite[Kim et al.][submitted]{}.
Optimistically speaking, {as more \textit{JWST} data become available, the evolutionary properties of different populations of galaxies will be easily revealed by the multi-band analysis of photometric/spectroscopic redshift information.}

In Fig. \ref{LSS}, we show the field-to-average fluctuations due to the large scale structures in 7.7, 10, and 15 $\mu$m compared with the source counts from the Stephen’s Quintet fields of \citet{Ling2022}. 
Fig. \ref{LSS} shows the ratio between the average of each bin and source counts with bin size $=0.5$ dex. 
The derived field-to-average fluctuation is majorly between 30\% to 180\%. 
Some of the bins have no sources in them thus yielding 0\% in the field-to-average ratio. 
The error of these fluctuations is calculated from the error propagation of the Poisson noise of the source counts. 
{We observe a significant scatter in the ratio beyond the 1 sigma error tolerance, indicating the potential influence of LSS on the source counts.}
{Furthermore, it is clear that the sources in the background of Stephan's Quintet have a reduced number density, representing only 70\% of the average, while CEERS fields tend to have a higher density of sources.}
\section{Conclusions}\label{S:conc}
Using the deep MIR data from the \textit{JWST} CEERS fields, we {derived} the source counts in the six MIRI (7.7-21.0 $\mu$m) bands of the \textit{JWST}. The 80\% completeness limits are 0.25 (0.25), 0.63 (0.63), 1.26 (1.26), 2.0 (2.0), 5.0 (5.0), and 13 (16) $\mu$Jy, in F770W, F1000W, F1280W, F1500W, F1800W, and F2100W filters, respectively. These numbers are comparable to the expected flux limits given the \textit{JWST} sensitivity, illustrating their unprecedented depths in the MIR.
Compared with previous work, our results provide more than 2 orders of magnitude deeper extension of the source counts in 7 to 15 $\mu$m bands, and about 1 and 0.5 order extension in 18- and 21- $\mu$m, respectively.
Apart from the 10 $\mu$m, the other wavelengths show reasonable agreements with the extrapolation of the current models, which are primarily based on previous IR observations/studies (e.g., \textit{ISO}, \textit{Spitzer}, and \textit{AKARI}).

Our \textit{JWST} source counts at 12.8, 18, and 21 $\mu$m compared to \citet{Ling2022} are newly obtained in this work and provide us with the deepest counts to date.
Even though we have some deviations between the currently available models and the measurement at the F1000W (Section \ref{S:res}) in this work, it is clear that new observations with the MIRI/\textit{JWST} have broadened our horizons to much fainter infrared galaxies. {The development of improved models to physically interpret MIR source counts} across a wide range of observed fluxes is urgently needed. In our companion paper, we fit the newly obtained source counts to constrain the evolution of IR galaxy populations (Kim et al., submitted).

\section*{Acknowledgements}
{The authors would like to express their sincere gratitude to the referee, Prof. Chris Pearson, for many insightful comments and suggestions, which have greatly improved the quality and clarity of this manuscript.} TG acknowledges the support of the National Science and Technology Council of Taiwan through grants 108-2628-M-007-004-MY3 and 111-2123-M-001-008-.
TH acknowledges the support of the National Science and Technology Council of Taiwan through grants 110-2112-M-005-013-MY3, 110-2112-M-007-034-, and 111-2123-M-001-008-.
This work is based on observations made with the NASA/ESA/CSA James Webb Space Telescope. The data were obtained from the Mikulski Archive for Space Telescopes at the Space Telescope Science Institute, which is operated by the Association of Universities for Research in Astronomy, Inc., under NASA contract NAS 5-03127 for \textit{JWST}. These observations are associated with the program ERO. Also, we thank Jamie, Chang at the Department of Physics, NTHU for his profound editing skill for the beautiful Fig. \ref{fig:1}.

\section*{Data Availability}
Early Release Observations obtained by \textit{JWST} MIRI are publicly available at \url{https://www.stsci.edu/jwst/science-execution/approved-programs/webb-first-image-observations}.
Other data underlying this article will be shared upon reasonable request to the corresponding author.



\bibliographystyle{mnras}
\bibliography{jwstpaper} 

\begin{thebibliography}{}
\makeatletter
\relax
\def\mn@urlcharsother{\let\do\@makeother \do\$\do\&\do\#\do\^\do\_\do\%\do\~}
\def\mn@doi{\begingroup\mn@urlcharsother \@ifnextchar [ {\mn@doi@}
  {\mn@doi@[]}}
\def\mn@doi@[#1]#2{\def\@tempa{#1}\ifx\@tempa\@empty \href
  {http://dx.doi.org/#2} {doi:#2}\else \href {http://dx.doi.org/#2} {#1}\fi
  \endgroup}
\def\mn@eprint#1#2{\mn@eprint@#1:#2::\@nil}
\def\mn@eprint@arXiv#1{\href {http://arxiv.org/abs/#1} {{\tt arXiv:#1}}}
\def\mn@eprint@dblp#1{\href {http://dblp.uni-trier.de/rec/bibtex/#1.xml}
  {dblp:#1}}
\def\mn@eprint@#1:#2:#3:#4\@nil{\def\@tempa {#1}\def\@tempb {#2}\def\@tempc
  {#3}\ifx \@tempc \@empty \let \@tempc \@tempb \let \@tempb \@tempa \fi \ifx
  \@tempb \@empty \def\@tempb {arXiv}\fi \@ifundefined
  {mn@eprint@\@tempb}{\@tempb:\@tempc}{\expandafter \expandafter \csname
  mn@eprint@\@tempb\endcsname \expandafter{\@tempc}}}

\bibitem[\protect\citeauthoryear{{Ashby} et~al.,}{{Ashby}
  et~al.}{2015}]{Ashby2015}
{Ashby} M.~L.~N.,  et~al., 2015, \mn@doi [\apjs] {10.1088/0067-0049/218/2/33},
  \href {https://ui.adsabs.harvard.edu/abs/2015ApJS..218...33A} {218, 33}

\bibitem[\protect\citeauthoryear{{Bertin} \& {Arnouts}}{{Bertin} \&
  {Arnouts}}{1996}]{Bertin1996}
{Bertin} E.,  {Arnouts} S.,  1996, \mn@doi [\aaps] {10.1051/aas:1996164}, \href
  {https://ui.adsabs.harvard.edu/abs/1996A&AS..117..393B} {117, 393}

\bibitem[\protect\citeauthoryear{Bradley et~al.,}{Bradley
  et~al.}{2022}]{Larry2022}
Bradley L.,  et~al., 2022, astropy/photutils: 1.5.0,
  \mn@doi{10.5281/zenodo.6825092}, \url
  {https://doi.org/10.5281/zenodo.6825092}

\bibitem[\protect\citeauthoryear{{Caputi} et~al.}{{Caputi}
  et~al.}{2007}]{Caputi2007}
{Caputi} K.~I.,  et~al., 2007, \mn@doi [\apj] {10.1086/512667}, \href
  {http://adsabs.harvard.edu/abs/2007ApJ...660...97C} {660, 97}

\bibitem[\protect\citeauthoryear{{Cowley}, {Baugh}, {Cole}, {Frenk}  \&
  {Lacey}}{{Cowley} et~al.}{2018}]{Cowley2018}
{Cowley} W.~I.,  {Baugh} C.~M.,  {Cole} S.,  {Frenk} C.~S.,   {Lacey} C.~G.,
  2018, \mn@doi [\mnras] {10.1093/mnras/stx2897}, \href
  {https://ui.adsabs.harvard.edu/abs/2018MNRAS.474.2352C} {474, 2352}

\bibitem[\protect\citeauthoryear{{Davidge}, {Serjeant}, {Pearson}, {Matsuhara},
  {Wada}, {Dryer}  \& {Barrufet}}{{Davidge} et~al.}{2017}]{Davidge2017}
{Davidge} H.,  {Serjeant} S.,  {Pearson} C.,  {Matsuhara} H.,  {Wada} T.,
  {Dryer} B.,   {Barrufet} L.,  2017, \mn@doi [\mnras] {10.1093/mnras/stx1935},
  \href {https://ui.adsabs.harvard.edu/abs/2017MNRAS.472.4259D} {472, 4259}

\bibitem[\protect\citeauthoryear{Elbaz et~al.}{Elbaz et~al.}{1999}]{Elbaz1999}
Elbaz D.,  et~al., 1999, Astron. Astrophys., 351, L37

\bibitem[\protect\citeauthoryear{{Elbaz}, {Cesarsky}, {Chanial}, {Aussel},
  {Franceschini}, {Fadda}  \& {Chary}}{{Elbaz} et~al.}{2002}]{Elbaz2002}
{Elbaz} D.,  {Cesarsky} C.~J.,  {Chanial} P.,  {Aussel} H.,  {Franceschini} A.,
   {Fadda} D.,   {Chary} R.~R.,  2002, \mn@doi [\aap]
  {10.1051/0004-6361:20020106}, \href
  {https://ui.adsabs.harvard.edu/abs/2002A&A...384..848E} {384, 848}

\bibitem[\protect\citeauthoryear{{Fazio} et~al.,}{{Fazio}
  et~al.}{2004}]{Fazio2004}
{Fazio} G.~G.,  et~al., 2004, \mn@doi [\apjs] {10.1086/422585}, \href
  {https://ui.adsabs.harvard.edu/abs/2004ApJS..154...39F} {154, 39}

\bibitem[\protect\citeauthoryear{{Finkelstein} et~al.,}{{Finkelstein}
  et~al.}{2017}]{Finkelstein2017}
{Finkelstein} S.~L.,  et~al., 2017, {The Cosmic Evolution Early Release Science
  (CEERS) Survey}, JWST Proposal ID 1345. Cycle 0 Early Release Science

\bibitem[\protect\citeauthoryear{{Gardner} et~al.}{{Gardner}
  et~al.}{2006}]{Gardner2006}
{Gardner} J.~P.,  et~al., 2006, \mn@doi [\ssr] {10.1007/s11214-006-8315-7},
  \href {https://ui.adsabs.harvard.edu/abs/2006SSRv..123..485G} {123, 485}

\bibitem[\protect\citeauthoryear{{Gehrels}}{{Gehrels}}{1986}]{Gehrels1986}
{Gehrels} N.,  1986, \mn@doi [\apj] {10.1086/164079}, \href
  {https://ui.adsabs.harvard.edu/abs/1986ApJ...303..336G} {303, 336}

\bibitem[\protect\citeauthoryear{{Goto} et~al.}{{Goto} et~al.}{2010}]{Goto2010}
{Goto} T.,  et~al., 2010, \mn@doi [\aap] {10.1051/0004-6361/200913182}, \href
  {http://adsabs.harvard.edu/abs/2010A%26A...514A...6G} {514, A6}

\bibitem[\protect\citeauthoryear{{Gruppioni}, {Lari}, {Pozzi}, {Zamorani},
  {Franceschini}, {Oliver}, {Rowan-Robinson}  \& {Serjeant}}{{Gruppioni}
  et~al.}{2002}]{Gruppioni2002}
{Gruppioni} C.,  {Lari} C.,  {Pozzi} F.,  {Zamorani} G.,  {Franceschini} A.,
  {Oliver} S.,  {Rowan-Robinson} M.,   {Serjeant} S.,  2002, \mn@doi [\mnras]
  {10.1046/j.1365-8711.2002.05672.x}, \href
  {https://ui.adsabs.harvard.edu/abs/2002MNRAS.335..831G} {335, 831}

\bibitem[\protect\citeauthoryear{{Gruppioni} et~al.}{{Gruppioni}
  et~al.}{2010}]{Gruppioni2010}
{Gruppioni} C.,  et~al., 2010, \mn@doi [\aap] {10.1051/0004-6361/201014608},
  \href {https://ui.adsabs.harvard.edu/abs/2010A&A...518L..27G} {518, L27}

\bibitem[\protect\citeauthoryear{{Gruppioni}, {Pozzi}, {Zamorani}  \&
  {Vignali}}{{Gruppioni} et~al.}{2011}]{Gruppioni2011}
{Gruppioni} C.,  {Pozzi} F.,  {Zamorani} G.,   {Vignali} C.,  2011, \mn@doi
  [\mnras] {10.1111/j.1365-2966.2011.19006.x}, \href
  {https://ui.adsabs.harvard.edu/abs/2011MNRAS.416...70G} {416, 70}

\bibitem[\protect\citeauthoryear{{Kalirai}}{{Kalirai}}{2018}]{Kalirai2018}
{Kalirai} J.,  2018, \mn@doi [Contemporary Physics]
  {10.1080/00107514.2018.1467648}, \href
  {https://ui.adsabs.harvard.edu/abs/2018ConPh..59..251K} {59, 251}

\bibitem[\protect\citeauthoryear{{Kessler} et~al.}{{Kessler}
  et~al.}{1996}]{Kessler1996}
{Kessler} M.~F.,  et~al., 1996, \aap, \href
  {https://ui.adsabs.harvard.edu/abs/1996A&A...315L..27K} {315, L27}

\bibitem[\protect\citeauthoryear{Ling et~al.,}{Ling et~al.}{2022}]{Ling2022}
Ling C.-T.,  et~al., 2022, \mn@doi [Monthly Notices of the Royal Astronomical
  Society] {10.1093/mnras/stac2716}, 517, 853

\bibitem[\protect\citeauthoryear{{Murakami} et~al.}{{Murakami}
  et~al.}{2007}]{Murakami2007}
{Murakami} H.,  et~al., 2007, \mn@doi [\pasj] {10.1093/pasj/59.sp2.S369}, \href
  {https://ui.adsabs.harvard.edu/abs/2007PASJ...59S.369M} {59, S369}

\bibitem[\protect\citeauthoryear{{Neugebauer} et~al.}{{Neugebauer}
  et~al.}{1984}]{Neugebauer1984}
{Neugebauer} G.,  et~al., 1984, \mn@doi [\apjl] {10.1086/184209}, \href
  {https://ui.adsabs.harvard.edu/abs/1984ApJ...278L...1N} {278, L1}

\bibitem[\protect\citeauthoryear{{Papovich} et~al.,}{{Papovich}
  et~al.}{2004}]{Papovich2004}
{Papovich} C.,  et~al., 2004, \mn@doi [\apjs] {10.1086/422880}, \href
  {https://ui.adsabs.harvard.edu/abs/2004ApJS..154...70P} {154, 70}

\bibitem[\protect\citeauthoryear{{Pearson}}{{Pearson}}{2005}]{Pearson2005}
{Pearson} C.,  2005, \mn@doi [\mnras] {10.1111/j.1365-2966.2005.08861.x}, \href
  {https://ui.adsabs.harvard.edu/abs/2005MNRAS.358.1417P} {358, 1417}

\bibitem[\protect\citeauthoryear{{Pearson} et~al.}{{Pearson}
  et~al.}{2010}]{Pearson2010}
{Pearson} C.~P.,  et~al., 2010, \mn@doi [\aap] {10.1051/0004-6361/200913382},
  \href {https://ui.adsabs.harvard.edu/abs/2010A&A...514A...8P} {514, A8}

\bibitem[\protect\citeauthoryear{{Pearson} et~al.}{{Pearson}
  et~al.}{2014}]{Pearson2014}
{Pearson} C.~P.,  et~al., 2014, \mn@doi [\mnras] {10.1093/mnras/stu1472}, \href
  {https://ui.adsabs.harvard.edu/abs/2014MNRAS.444..846P} {444, 846}

\bibitem[\protect\citeauthoryear{{Pilbratt} et~al.}{{Pilbratt}
  et~al.}{2010}]{Pilbratt2010}
{Pilbratt} G.~L.,  et~al., 2010, \mn@doi [\aap] {10.1051/0004-6361/201014759},
  \href {https://ui.adsabs.harvard.edu/abs/2010A&A...518L...1P} {518, L1}

\bibitem[\protect\citeauthoryear{{Planck Collaboration} et~al.,}{{Planck
  Collaboration} et~al.}{2016}]{Planck2016}
{Planck Collaboration} et~al., 2016, \mn@doi [\aap]
  {10.1051/0004-6361/201525830}, \href
  {https://ui.adsabs.harvard.edu/abs/2016A&A...594A..13P} {594, A13}

\bibitem[\protect\citeauthoryear{{Rocca-Volmerange}, {de Lapparent}, {Seymour}
  \& {Fioc}}{{Rocca-Volmerange} et~al.}{2007}]{Rocca2007}
{Rocca-Volmerange} B.,  {de Lapparent} V.,  {Seymour} N.,   {Fioc} M.,  2007,
  \mn@doi [\aap] {10.1051/0004-6361:20065217}, \href
  {https://ui.adsabs.harvard.edu/abs/2007A&A...475..801R} {475, 801}

\bibitem[\protect\citeauthoryear{{Rowan-Robinson} et~al.}{{Rowan-Robinson}
  et~al.}{1997}]{Rowan-Robinson1997}
{Rowan-Robinson} M.,  et~al., 1997, \mn@doi [\mnras] {10.1093/mnras/289.2.490},
  \href {https://ui.adsabs.harvard.edu/abs/1997MNRAS.289..490R} {289, 490}

\bibitem[\protect\citeauthoryear{{Saunders}, {Rowan-Robinson}, {Lawrence},
  {Efstathiou}, {Kaiser}, {Ellis}  \& {Frenk}}{{Saunders}
  et~al.}{1990}]{Saunders1990}
{Saunders} W.,  {Rowan-Robinson} M.,  {Lawrence} A.,  {Efstathiou} G.,
  {Kaiser} N.,  {Ellis} R.~S.,   {Frenk} C.~S.,  1990, \mn@doi [\mnras]
  {10.1093/mnras/242.3.318}, \href
  {http://adsabs.harvard.edu/abs/1990MNRAS.242..318S} {242, 318}

\bibitem[\protect\citeauthoryear{{Takagi} et~al.}{{Takagi}
  et~al.}{2012}]{Takagi2012}
{Takagi} T.,  et~al., 2012, \mn@doi [\aap] {10.1051/0004-6361/201117759}, \href
  {https://ui.adsabs.harvard.edu/abs/2012A&A...537A..24T} {537, A24}

\bibitem[\protect\citeauthoryear{{Wada} et~al.}{{Wada} et~al.}{2008}]{Wada2008}
{Wada} T.,  et~al., 2008, \mn@doi [\pasj] {10.1093/pasj/60.sp2.S517}, \href
  {https://ui.adsabs.harvard.edu/abs/2008PASJ...60S.517W} {60, S517}

\bibitem[\protect\citeauthoryear{{Werner} et~al.}{{Werner}
  et~al.}{2004}]{Werner2004}
{Werner} M.~W.,  et~al., 2004, \mn@doi [\apjs] {10.1086/422992}, \href
  {https://ui.adsabs.harvard.edu/abs/2004ApJS..154....1W} {154, 1}

\makeatother
\end{thebibliography}




%

\appendix 
\section{Appendix}
\label{Appendix}
Here we put the randomly selected example sources from each band of the \textit{JWST} MIRI filter to demonstrate differences in flux levels.
The fluxes of the objects in the first row are roughly at the o001 (o002) 80\% completeness: {0.25 (0.25), 0.63 (0.63), 1.26 (1.26), 2.0 (2.0), 5.0 (5.0), and 13 (16)} $\mu$Jy for different filters, respectively. While the images in the second row are 10 to 1000 times brighter.

\begin{figure}
    \centering
    \includegraphics[width=\columnwidth]{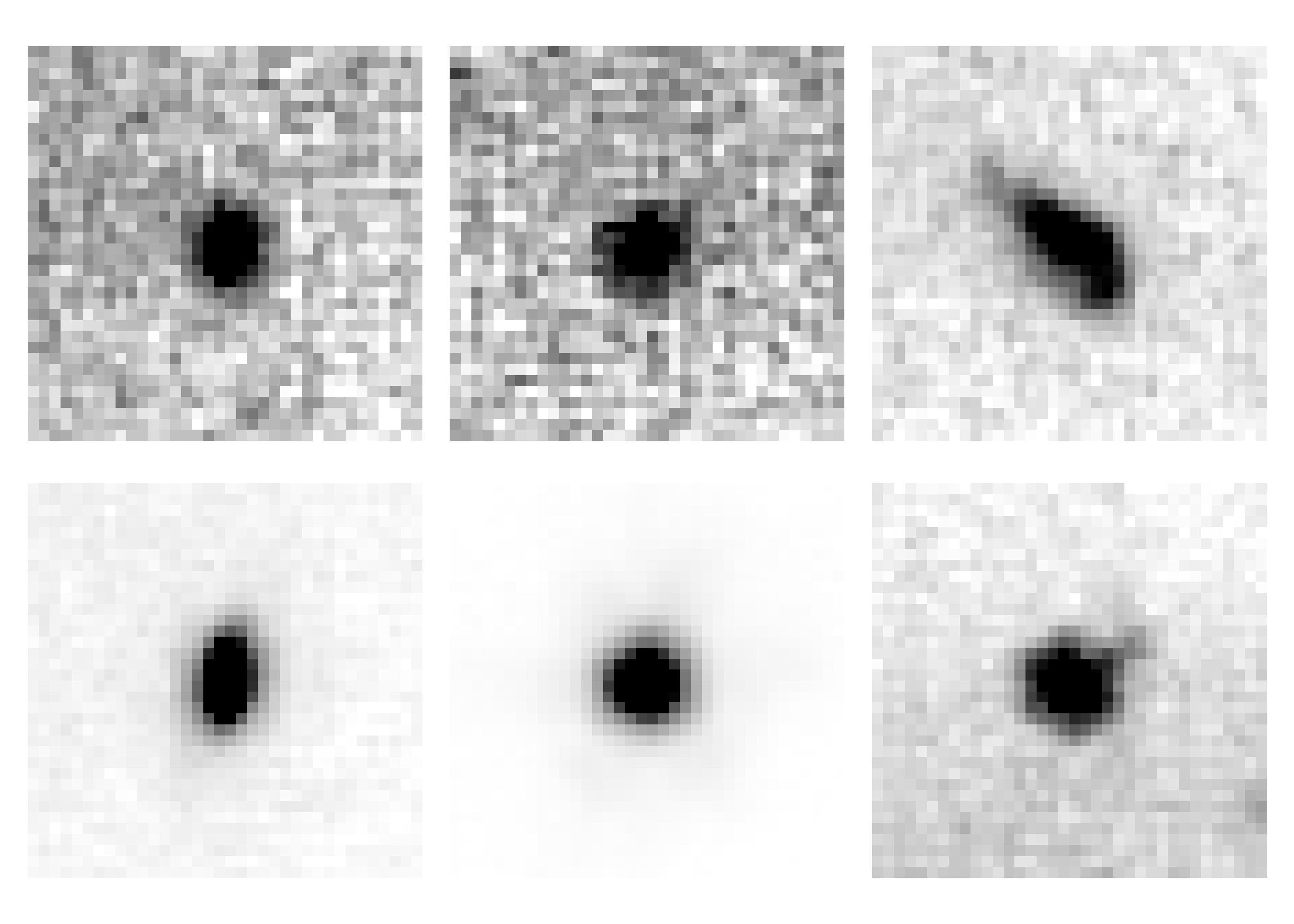}
    \caption{Examples of six detected sources in field o001 at the 7.7-$\mu$m band (F770W). The fluxes from objects in the first row sit roughly at 80\% completeness while they are 10 to 1000 times brighter in the second row. All images have a field-of-view of 2$\times$2 arcsec$^{2}$.}
    \label{o001f770source}
\end{figure}

\begin{figure}
    \centering
     \includegraphics[width=\columnwidth]{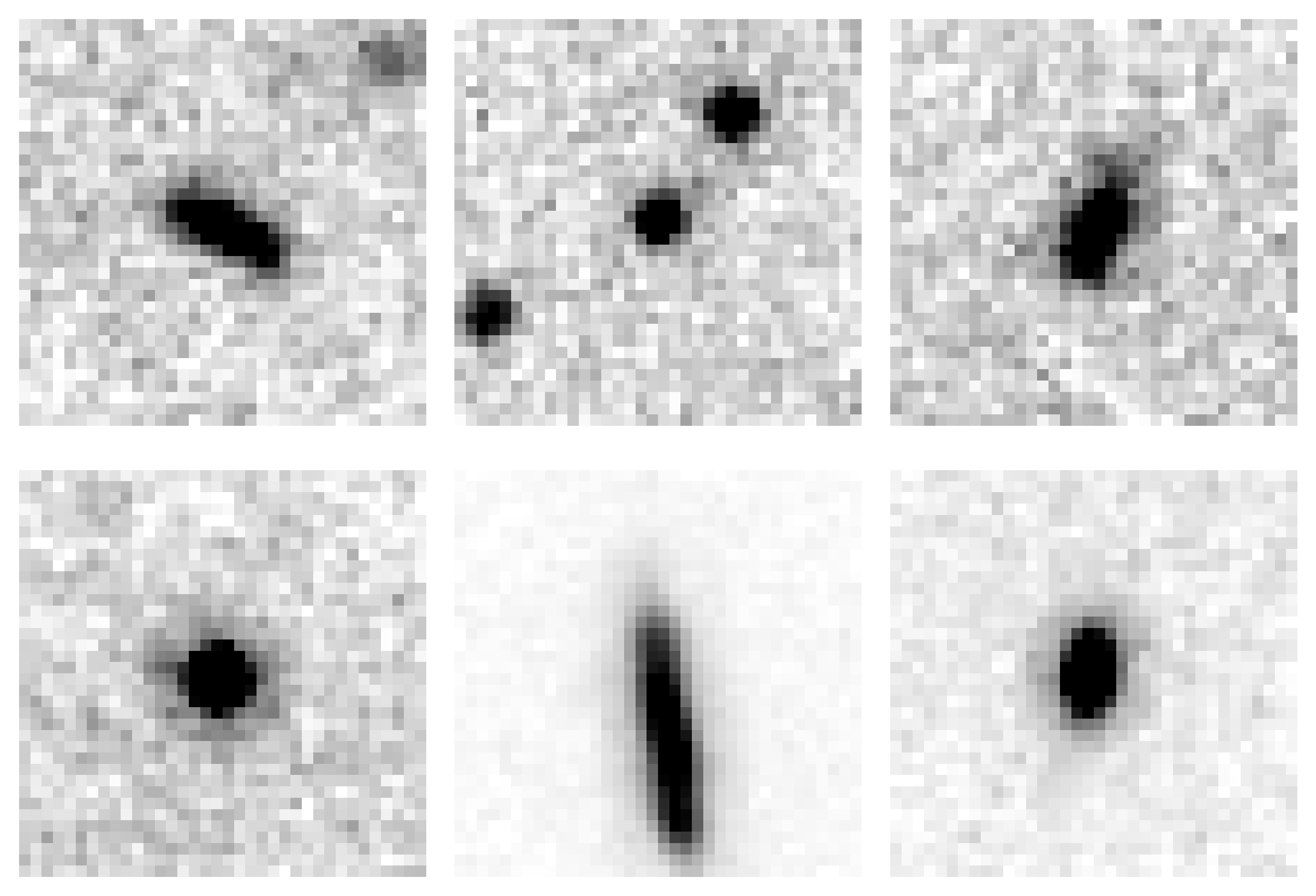}
    \caption{Same figure as Fig. \ref{o001f770source} but for 10.0-$\mu$m band (F1000W).}
    \label{o001f1000source}
\end{figure}

\begin{figure}
    \centering
     \includegraphics[width=\columnwidth]{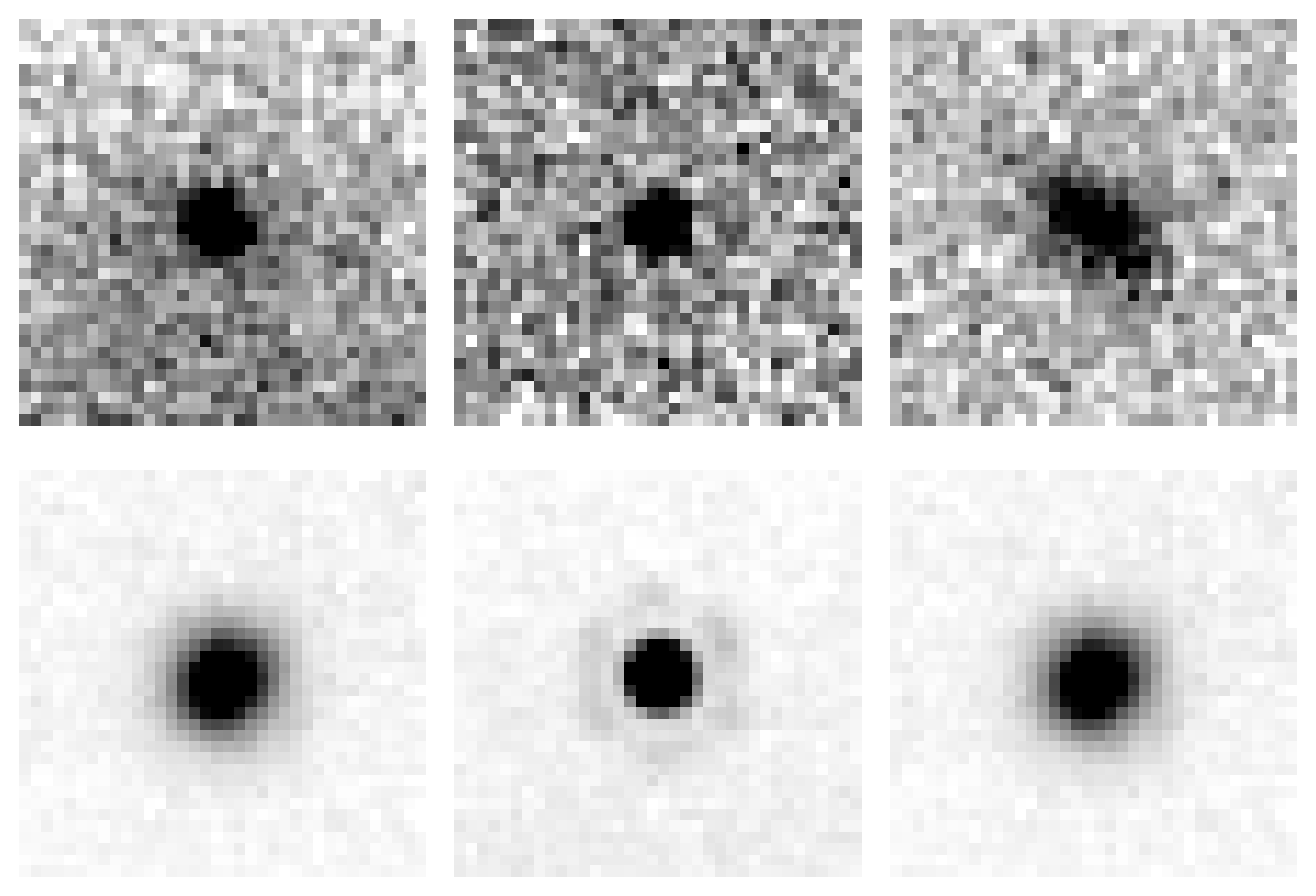}
    \caption{Same figure as Fig. \ref{o001f770source} but for 12.8-$\mu$m band (F1280W).}
    \label{o001f1280source}
\end{figure}

\begin{figure}
    \centering
     \includegraphics[width=\columnwidth]{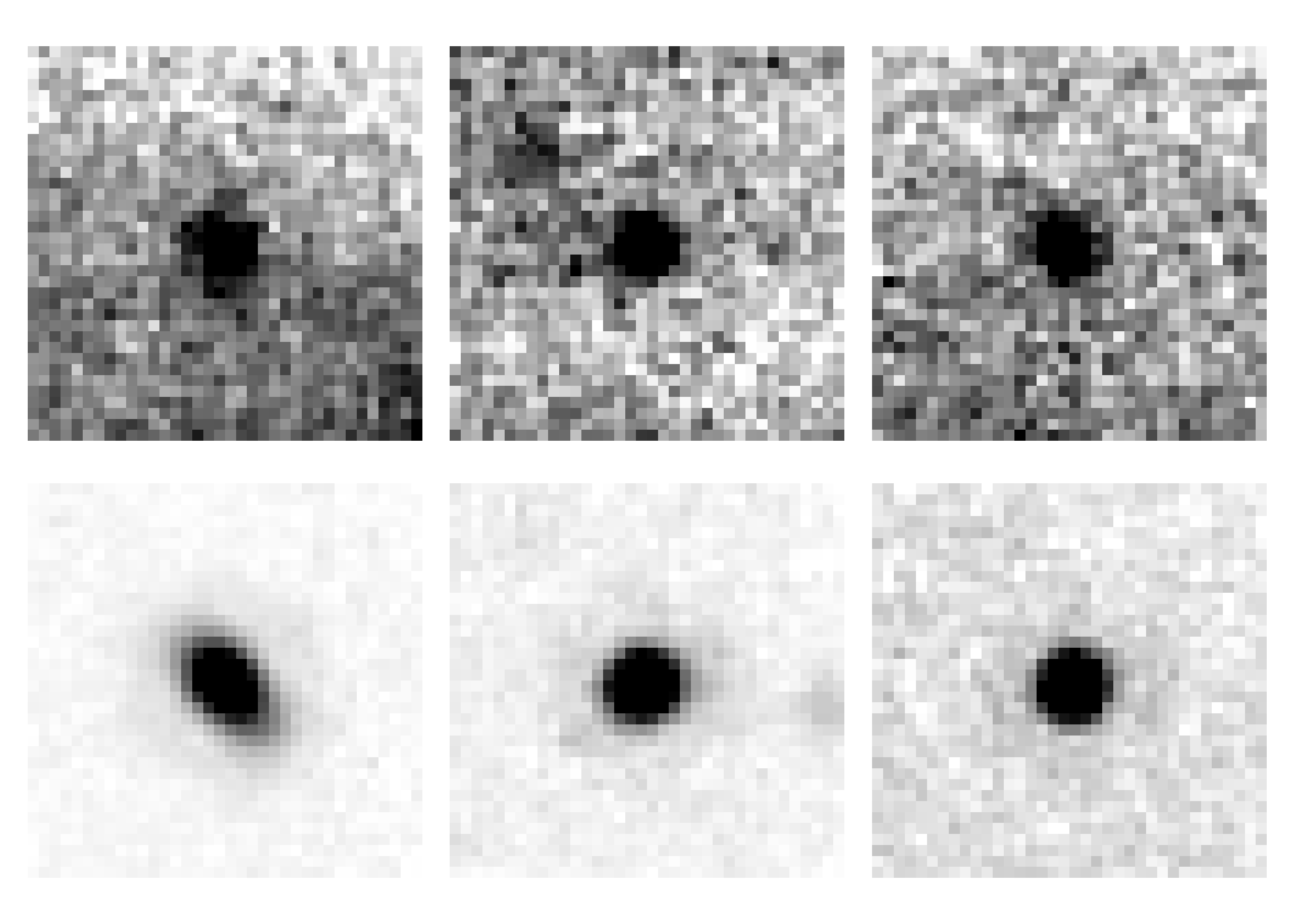}
    \caption{Same figure as Fig. \ref{o001f770source} but for 15.0-$\mu$m band (F1500W).}
    \label{o001f1500source}
\end{figure}

\begin{figure}
    \centering
     \includegraphics[width=\columnwidth]{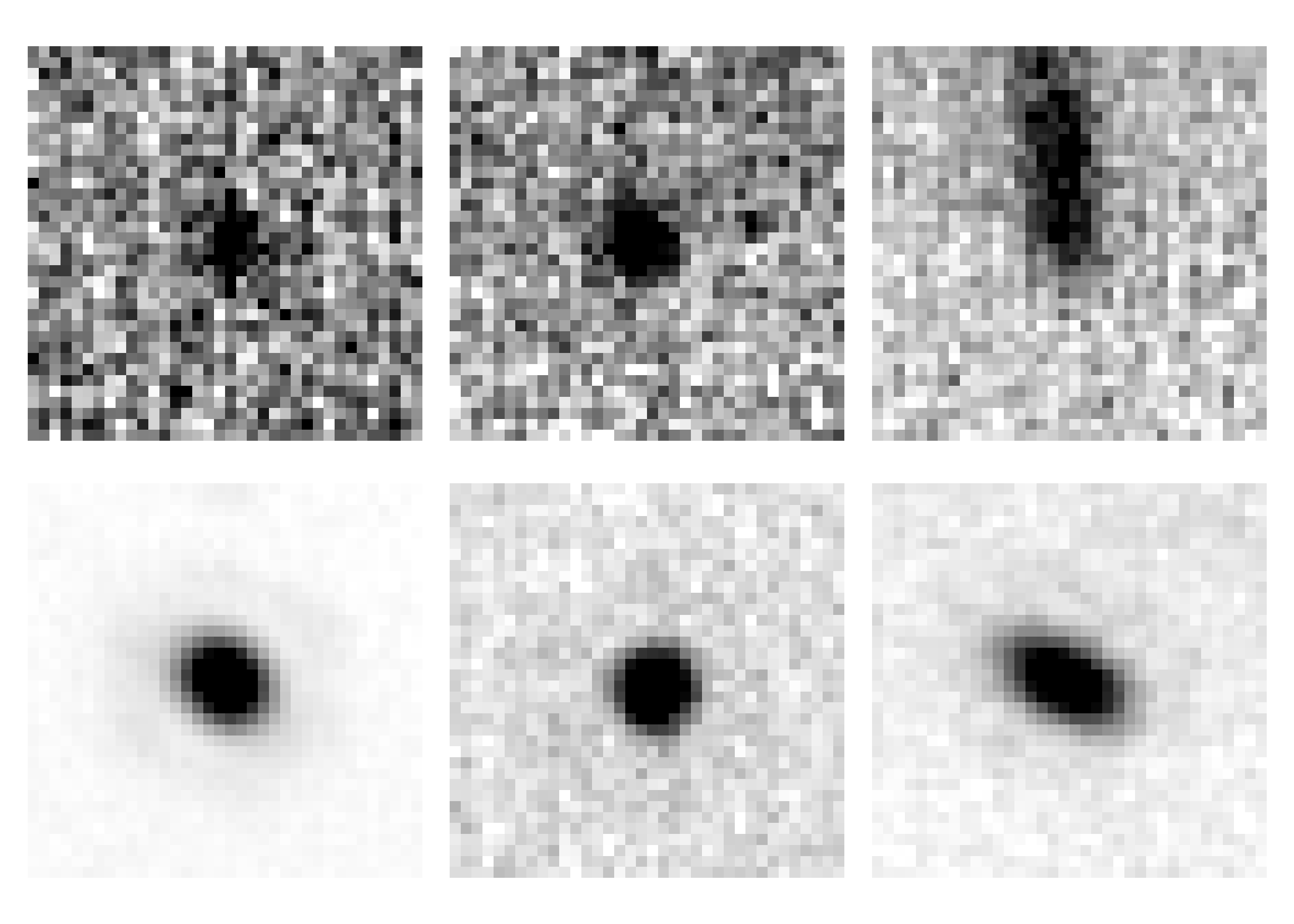}
    \caption{Same figure as Fig. \ref{o001f770source} but for 18.0-$\mu$m band (F1800W).}
    \label{o001f1800source}
\end{figure}

\begin{figure}
    \centering
     \includegraphics[width=\columnwidth]{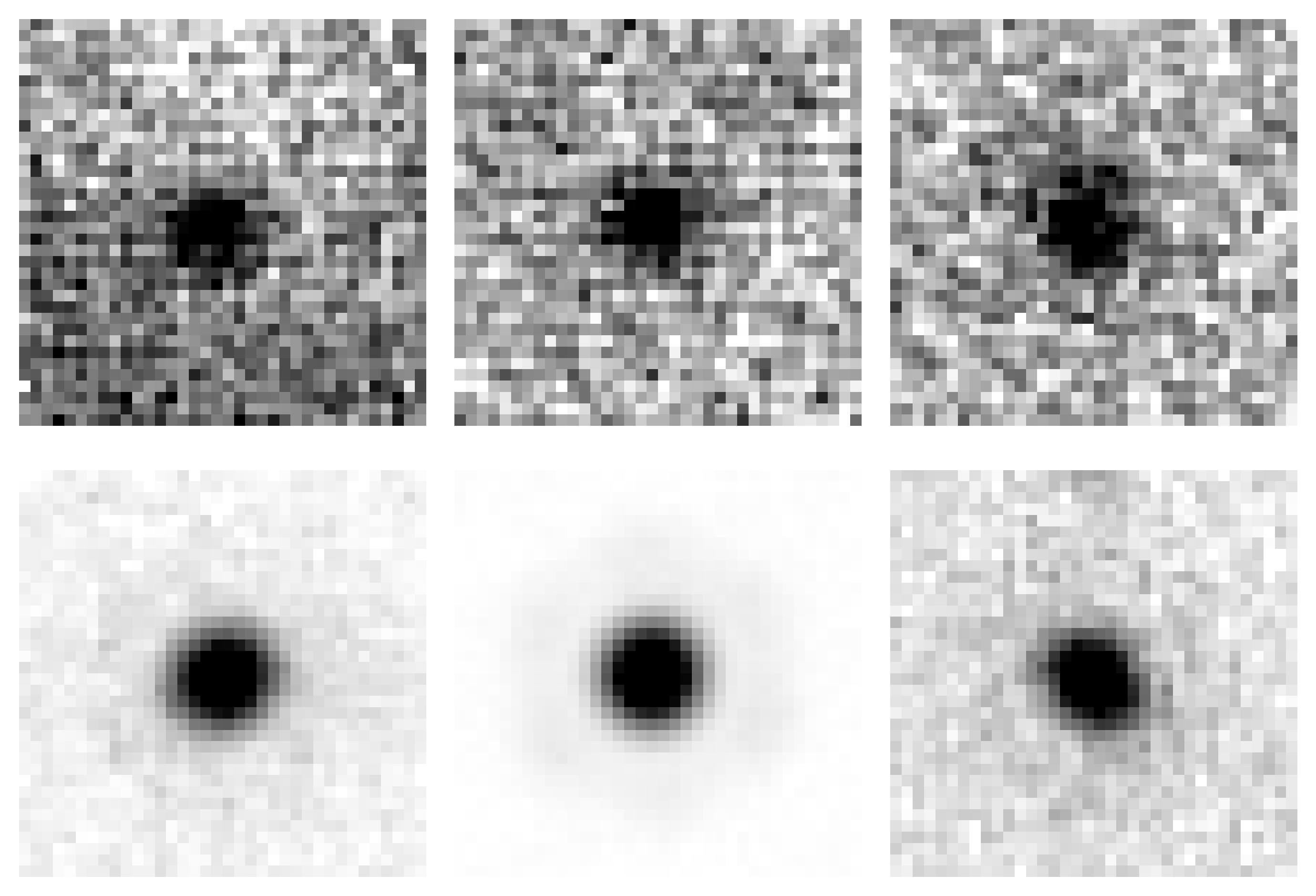}
    \caption{Same figure as Fig. \ref{o001f770source} but for 21.0-$\mu$m band (F2100W).}
    \label{o001f2100source}
\end{figure}


\bsp	
\label{lastpage} 
\end{document}